\documentclass[preprint,11.5pt]{elsarticle}

\journal{Computer Physics Communications}

\usepackage{graphicx}

\usepackage{xcolor}

\usepackage{amsmath}
\usepackage{amssymb}

\usepackage{bm}
\usepackage{booktabs}

\usepackage{multirow}
\usepackage{dcolumn}

\usepackage{adjustbox}
\usepackage{array} 
\usepackage{float}
\usepackage[section]{placeins}

\newcommand{\dash}{\multicolumn{1}{c}{\textcolor{black}{--}}}

\usepackage{caption}
\usepackage{hyperref}
\hypersetup{
    colorlinks=true,
    linkcolor=blue,
    citecolor=blue,
    filecolor=blue,
    urlcolor=blue,
    pdftitle={SEP TMDC monolayer},
    pdfauthor={Raj Kumar Paudel},     
}

\begin{document}

\pagestyle{empty}

\begin{frontmatter}

\title{Semi-empirical Pseudopotential Method for Monolayer Transition Metal Dichalcogenides}

\author[label1,label2]{Raj Kumar Paudel}
\ead{rajupdl6@phys.ncku.edu.tw}

\author[label3]{Chung-Yuan Ren}

\author[label1,label2]{Yia-Chung Chang}
\ead{yiachang@gate.sinica.edu.tw (corresponding author)}

\address[label1]{Research Center for Applied Sciences, Academia Sinica, Taipei 11529, Taiwan}
\address[label2]{Department of Physics, National Cheng-Kung University, Tainan 701, Taiwan}
\address[label3]{Department of Physics, National Kaohsiung Normal University, Kaohsiung 824, Taiwan}

\begin{abstract}
We present a semi-empirical pseudopotential method for accurately computing the
band structures and Bloch states of monolayer transition metal dichalcogenides
(TMDCs), including MoS$_2$, MoSe$_2$, WS$_2$, and WSe$_2$. Our approach combines
local and nonlocal pseudopotentials, carefully fitted to reproduce fully
self-consistent density-functional theory results while using only a minimal set
of empirical parameters. By expressing the total potential as a sum of a few
separable components, we achieve both accuracy and computational efficiency.
The transferability of the monolayer-fitted pseudopotentials is assessed through
a direct application to bilayer TMDCs without additional refitting, where good
agreement with self-consistent DFT band structures is obtained near the band
edges. The resulting framework provides an efficient and flexible platform for
band-structure and Bloch-state calculations in TMDC-based low-dimensional
materials.
\end{abstract}


\end{frontmatter}

%\linenumbers

\section{Introduction}

Layered transition metal dichalcogenides (TMDCs) are a class of two-dimensional (2D) materials in which covalently bound layers are stacked together by relatively weak van der Waals forces, fundamentally different from their bulk counterparts. The discovery of graphene in 2004 \cite{novoselov2004electric} ignited extensive research into 2D materials due to its exceptional properties; however, its lack of a band gap limits its applications. This has caused a shift in focus to two-dimensional TMDCs \cite{mak2010atomically,wang2012electronics,mattheiss1973band}, such as MoS$_2$, MoSe$_2$, MoTe$_2$, WS$_2$, WSe$_2$, and WTe$_2$, which offer various electronic properties, ranging from semiconducting behavior to superconductivity \cite{qi2016superconductivity}. TMDCs with direct band gaps \cite{mak2010atomically} are promising for use in next-generation optoelectronics \cite{liu2014optical,wang2012electronics,qiu2013optical}, mechanical \cite{johari2012tuning,shi2018mechanical}, spintronics \cite{zhu2011giant,yun2022escalating}, and electrocatalysis \cite{li2018metallic} systems for energy storage and conversion. These materials also exhibit unique topological properties \cite{wu2019topological}, making them promising for novel electronic and quantum-computing applications. Recent advances include the successful fabrication of high-performance field effect transistors \cite{chen2020environmental}, phototransistors \cite{li2014metal}, and gas sensors \cite{ko2016improvement}, which highlight the potential of TMDCs in advanced technological applications and have attracted significant attention from both the scientific and the industrial communities.

First-principles density functional theory (DFT) \cite{hohenberg1964inhomogeneous,kohn1965self,martin2020electronic,ren2022density,ren2023density} combined with pseudopotential schemes \cite{herring1940new,phillips1959new,chelikowsky1976nonlocal,wang1995local,wang1996pseudopotential,bester2008electronic} remains the workhorse for predicting electronic properties. However, for 2D materials like TMDCs, conventional approaches using three-dimensional plane waves (PWs) face challenges due to the need for large vacuum spaces in supercell calculations to mimic three-dimensional periodicity. Earlier work by Li et al.\ \cite{li1994electronic,chang1996planar} indicated that a planar-based approach that combines plane waves in the periodic directions (x-y plane) with Gaussian functions in the non-periodic z direction is capable of accurate calculations of total energy and work function for isolated slabs. Ren et al.\ \cite{ren2015mixed,ren2022density} refined the above method using a mixed-basis approach that replaces Gaussian functions in the z direction with B-spline functions \cite{deBoor1987practical,bachau2001applications}. The use of B-splines is advantageous in accurately representing both fast-varying and slow-varying electronic wavefunctions. The mixed-basis approach offers advantages in preserving the layer-like geometry and reducing computational overhead in solving the Kohn–Sham equations. It also automatically avoids the vacuum region used in 3D plane-wave calculations to prevent interactions between periodic images.

The primary motivation for developing the semi-empirical Pseudopotential (SEP) Method \cite{wang1996pseudopotential,paudel2023semi} lies in the computational efficiency achieved by avoiding the self-consistent density optimization required in density functional theory (DFT). This approach is particularly beneficial for nanoscale structures comprising thousands to hundreds of thousands of atoms. Initially, empirical pseudopotentials defined in reciprocal space \cite{cohen1966band,chelikowsky1976nonlocal,pandey1974nonlocal} were fitted to match experimentally determined energy levels, allowing accurate predictions of band structures and optical properties with minimal Fourier components. However, these pseudopotentials lacked transferability between different structures. This issue was addressed by developing continuous pseudopotentials in real space, enabling accurate calculations of the electronic, optical, and transport properties in nanostructures using advanced computational methods \cite{wang1995local,fu1997local,bester2008electronic,molina2012semiempirical}.

Recent machine learning (ML) has transformed empirical pseudopotentials \cite{kim2024transferable}, providing new avenues for improving their transferability and accuracy. Momentum-range-separated, rotation-covariant descriptors capture local symmetry and bond-directional information on anisotropic solids. The ``DeepPseudopot'' model \cite{lin2025deep} integrates neural networks for local pseudopotentials with parameterized non-local spin-orbit coupling factors. However, these ML-driven methods have focused only on bulk systems.

In our previous research \cite{paudel2023semi}, SEP was introduced for 2D systems such as graphene and graphene nanoribbons using a mixed basis approach. Using the semi-empirical pseudopotential (SEP) with just a few parameters, we successfully replicated the complete band structure of graphene. The SEP can also be used to model armchair graphene nanoribbons (aGNRs) by adding the edge-induced correction potential (EICP). The SEP and EICP are applicable for aGNRs of any size, demonstrating the transferability of the SEP for application in large-scale aGNRs. In the present work, we extend the SEP to two‐dimensional TMDC monolayers, where the more complex crystal structure (with three atoms per unit cell) poses additional challenges.

The remainder of this paper is organized as follows. Section 2 presents the SEP formalism and details of our mixed-basis construction. Section 3 applies the method to monolayer TMDCs and examine the transferability to bilayer systems, comparing the electronic properties obtained using SEP and DFT. Section 4 concludes with summary and an outlook for future extensions.

\section{Methodology}

\subsection{Computational Details}

Electronic-structure calculations were performed within density functional theory (DFT) using the Vanderbilt ultrasoft pseudopotential (USPP) method~\cite{vanderbilt1990soft} to describe the ionic cores. A mixed-basis approach~\cite{ren2015mixed,ren2022density,paudel2023semi}, combining plane waves in the periodic directions and B-spline functions in the non-periodic direction, was employed. This methodology has been previously validated for low-dimensional systems through direct comparison with plane-wave calculations using VASP~\cite{kresse1996efficiency}. The exchange--correlation effects were treated within the generalized gradient approximation using the Perdew--Burke--Ernzerhof (PBE) functional~\cite{perdew1996generalized}.

Eigenvalue problems were solved using a conjugate-gradient algorithm~\cite{shewchuk1994introduction}, and structural relaxations were carried out using the Broyden--Fletcher--Goldfarb--Shanno (BFGS) algorithm~\cite{QE-2009}. The in-plane lattice vectors form a hexagonal geometry, while the out-of-plane direction was represented using real-space B-splines~\cite{deBoor1987practical}. Real-space integrations were performed on a uniform FFT grid of $24 \times 24$ grid points in the in-plane directions. All structures were fully relaxed until the total energy converged within $10^{-7}$~eV. Numerical convergence was ensured using a plane-wave energy cutoff of 30~Ry together with 25 B-splines distributed over a range of $4a$, where $a$ is the lattice constant. A $6\times6\times1$ Monkhorst--Pack $k$-point mesh was used. For  transition metal dichalcogenides containing $d$ electrons of Mo and W atoms, non-local core corrections were included. Self-consistent calculations were performed until the variation in the effective potential was less than $10^{-7}$~Ry. For bilayer TDMCs,  we use 43 B-spline basis functions distributed over a domain of $6a$.

\subsection{Atomic Structure and Basis Functions}
\label{sec:atomic-basis}

\begin{figure}[h]
  \centering
  \includegraphics[width=\linewidth]{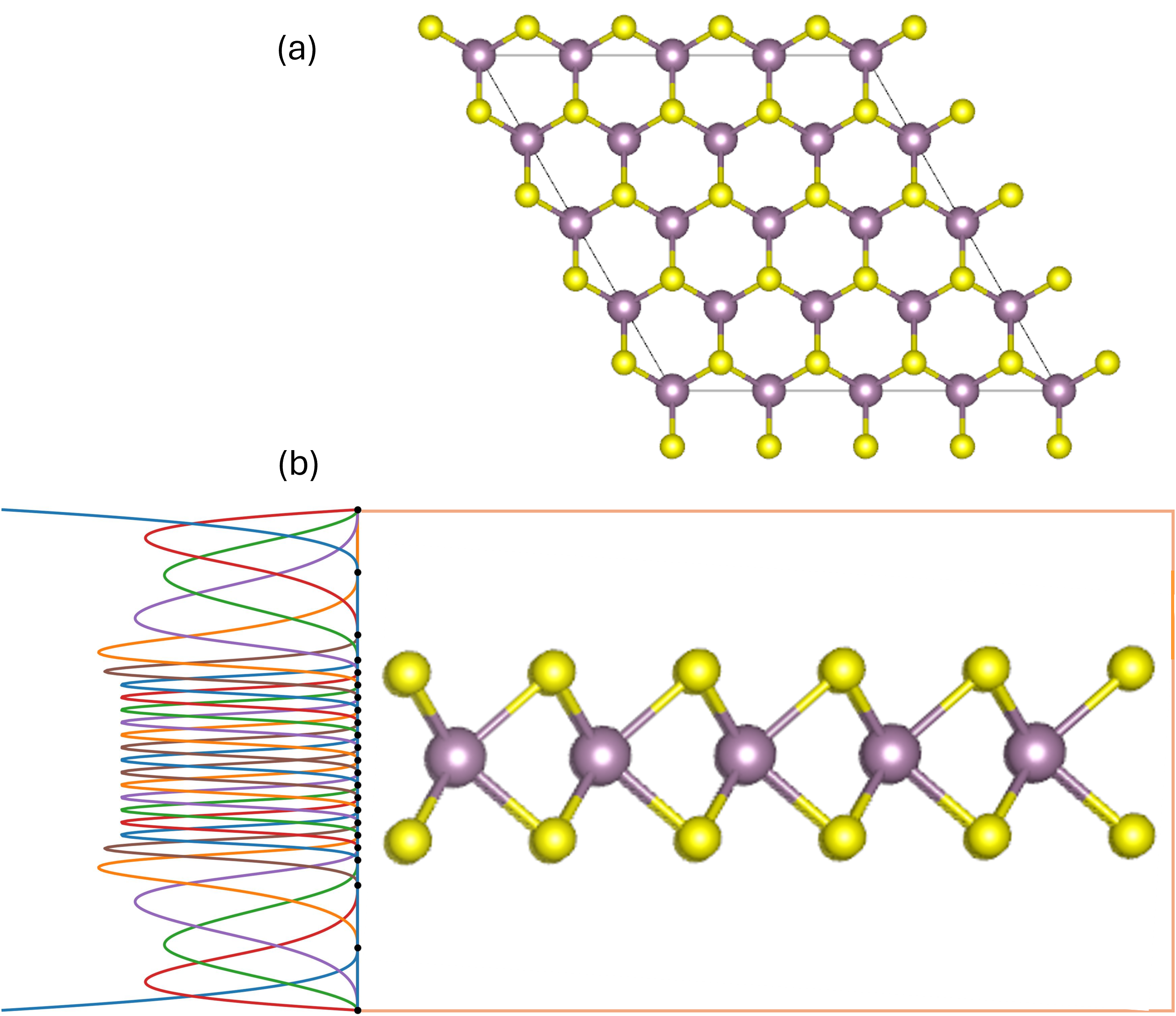}
  \caption{(a) Top view of a TMDC monolayer (M = metal, X = chalcogen). (b) Side view showing the same atomic layers (right) alongside the cubic B-spline basis functions (left) defined on a 29-point knot grid (black dots) spanning $4a$ with spline order $\kappa=4$. The 29-point knot points (in unit of $a$) are:\{0,0,0,0,0.5,1,1.2,1.3,1.4,1.5,1.6,1.7,1.8,1.9,2,2.1,2.2,2.3,2.4,2.5,2.6,2.7,2.8,3,3.5,4,4,4,4\}. The colored curves are the resulting k = 4 (cubic) B-spline basis functions; this knot sequence generates 25 splines.}
  \label{fig:struct_bspline}
\end{figure}

We consider monolayer transition‐metal dichalcogenides (TMDCs) in their hexagonal (2H) phase. Figure~\ref{fig:struct_bspline}(a) shows the top view of the atomic lattice, where the large (brown) spheres denote the metal atoms (M) and the smaller (yellow) spheres denote the chalcogen atoms (X). In the side view [Fig.~\ref{fig:struct_bspline}(b)], the atomic planes are aligned with a set of cubic B‐spline basis functions along the $z$direction.

B‐splines of order~$\kappa$ (polynomial degree $\kappa-1$) are defined recursively on the knot sequence $\{t_i\}$ \cite{deBoor1987practical,bachau2001applications}:
\begin{equation}
  B_{i,\kappa}(z) =
    \frac{z - t_{i}}{t_{i+\kappa-1} - t_{i}}\,B_{i,\kappa-1}(z)
    +
    \frac{t_{i+\kappa} - z}{t_{i+\kappa} - t_{i+1}}\,B_{i+1,\kappa-1}(z)
\end{equation}
with the zeroth‐order functions
\begin{equation}
  B_{i,1}(z) =
  \begin{cases}
    1, & t_i \le z < t_{i+1}\\
    0, & \text{otherwise}
  \end{cases}
\end{equation}

Following the mixed‐basis pseudopotential approach \cite{ren2015mixed,ren2022density,paudel2023semi}, the Bloch states of the monolayer TMDCs considered are expanded within a set of basis functions which are products of B-splines in the $z$ direction and 2D plane waves in the $x-y$ plane labeled by the 2D reciprocal lattice vectors $\mathbf{G}$. We denote the basis set as $\{B_i(z),\mathbf{G}\}$. The effectiveness of B-splines lies in their flexibility through adjustable knot positions, and they are straightforward to evaluate and differentiate.

\subsection{Construction of SEP from USPP}
\label{sec:level2}

In the 2H phase, each transition-metal atom is coordinated by six chalcogen atoms in a trigonal prismatic geometry. A key feature of these materials is the transition of the fundamental band gap from an indirect gap in the bulk to a direct gap in the monolayer limit, making them highly attractive for electronic and optoelectronic applications.
In the monolayer form, the structure belongs to the space group $P\overline{6}m2$ (No.~187), corresponding to the $D_{3h}$ point group symmetry \cite{gorelik20213d}. The transition-metal atom (Mo or W) is located at $(0,0,0)$, while the chalcogen atoms (S or Se) are positioned at $\mathbf{a}_1/3+2\mathbf{a}_2/3\pm u \mathbf{a}_3$. Here, $\mathbf{a}_1$, $\mathbf{a}_2$, and $\mathbf{a}_3$ are primitive lattice vectors. The internal structural parameter $u$ is determined by performing full structural relaxation via total-energy minimization.

We conduct systematic structural optimization for the four monolayer TMDC compounds MX$_2$ considered. The primitive lattice vectors are defined as $\mathbf{a}_1 = (1/2,-\sqrt{3}/2) a$ and $\mathbf{a}_2 = (  1/2,\sqrt{3}/2 ) a$. The basis vectors in reciprocal space are $\mathbf{b}_1 = \frac{2\pi}{a} ( 1, -1/\sqrt{3} )$ and $\mathbf{b}_2 = \frac{2\pi}{a} (1, 1/\sqrt{3} )$. Here, $a$ is the lattice constant.

Our SEP methodology starts by solving the Kohn-Sham equations \cite{kohn1965self} within the mixed-basis approach \cite{ren2015mixed}. This equation comprehensively accounts for kinetic energy contributions alongside potential energy terms arising from both local and non-local pseudopotential components. We extract the local portion of the self-consistent effective potential from the DFT calculation in the mixed basis. In density functional theory (DFT), the electronic states of a solid are obtained by self-consistently solving the single-particle effective Schrödinger equations:
\begin{equation}\label{eq:SWEdft}
\left\{ -{\nabla^{2}} + V_{\text{eff}}(\mathbf{r}) + \hat{V}_{\text{nloc}}(\mathbf{r}) \right\} \psi_{i}(\mathbf{r}) = \varepsilon_{i} \psi_{i}(\mathbf{r})
\end{equation}
where the $-{\nabla^{2}}$ term describes the kinetic energy operator, $V_{\text{eff}}(\mathbf{r})$ denotes the net effective local potential, and $\hat{V}_{\text{nloc}}(\mathbf{r})$ represents the nonlocal pseudopotential.

The net effective potential $V_{\text{eff}}(\mathbf{r})$ is given by:
\begin{equation}\label{eq:Veff}
V_{\text{eff}}(\mathbf{r}) = V_{\text{ion}}(\mathbf{r}) + V_{\text{hxc}}[\rho(\mathbf{r})]
\end{equation}
where $V_{\text{ion}}(\mathbf{r})$ represents the local ionic pseudopotential and $V_{\text{hxc}}[\rho(\mathbf{r})]$ accounts for the Hartree and exchange-correlation contributions:
\begin{equation}\label{eq:Vhxc}
V_{\text{hxc}}[\rho(\mathbf{r})] = \int \frac{\rho(\mathbf{r}')}{|\mathbf{r} - \mathbf{r}'|} \, \mathrm{d}^3\mathbf{r}' + V_{\text{xc}}[\rho(\mathbf{r})]
\end{equation}

Here $\rho(\mathbf{r}) = \sum_{i} |\psi_i(\mathbf{r})|^2$ denotes the charge density of electrons in all occupied states. $V_{\text{hxc}}[\rho(\mathbf{r})]$ is specific to the system and must be determined self-consistently.

\subsubsection{Local Ionic Potential}

The local ionic potential $V_{ion}(\mathbf{r})$ is taken from the ultra-soft pseudopotential (USPP) \cite{vanderbilt1990soft} of the relevant atoms. In the USPP method, the local ionic potential $V_{\text{ion}}(\mathbf{r})$ is constructed to accurately reproduce the scattering properties of valence electrons in each atom while ensuring smoothness and transferability between different environments. It can be expressed as:
\begin{equation}\label{eq:vlocdft}
V_{ion}(\mathbf{r})=\sum_{\sigma,\mathbf{R}}[ V_{c, LR}^{\sigma}(\mathbf{r-R}) + V_{c,SR}^{\sigma}(\mathbf{r-R})]
\end{equation}
where $\sigma$ labels the atoms within each unit cell of the crystal, $\mathbf{R}$ labels the lattice sites in the crystal, $V_{c, LR}^{\sigma}(\mathbf{r})$ represents the long-range Coulomb potential due to the ionic core that decreases slowly with distance, and $V_{c, SR}^{\sigma}(\mathbf{r})$ is a short-range term that contains the finer details of the core potential close to the nucleus.

The first term in Eq.~(\ref{eq:vlocdft}) represents the potential from an auxiliary charge distribution, calculated analytically as:
\begin{align}
    V_{c,LR}^{\sigma}(\mathbf{r}) &= \int \frac{\rho_{c}^{\sigma}(\mathbf{r}')}{|\mathbf{r} - \mathbf{r}'|} \, d\mathbf{r}' \nonumber \\
    &= \sum_{\mathbf{q}, \mathbf{q}'} \int e^{i\mathbf{q} \cdot (\mathbf{r} - \mathbf{r}')} e^{i\mathbf{q}' \cdot \mathbf{r}'} \frac{8\pi}{q^2} \widetilde{\rho}_{c}^{\sigma}(\mathbf{q}') \, d\mathbf{r}' \nonumber \\
    &= \sum_{\mathbf{q}} \int e^{i\mathbf{q} \cdot \mathbf{r}} \frac{8\pi}{q^2} \widetilde\rho_{c}^{\sigma}(\mathbf{q}) \nonumber \\
    &= 8\pi \sum_{\mathbf{q}_{\parallel}, q_{z}} \int e^{i\mathbf{q} \cdot \mathbf{r}} \frac{1}{q_{\parallel}^{2} + q_{z}^{2}} \widetilde{\rho}_{c}^{\sigma}(\mathbf{q}_{\parallel} + q_{z} \widehat{\mathbf{z}}) \label{eq:Vclr_final}
\end{align}
where $\widetilde\rho_{c}^{\sigma}(\mathbf{q}) = \frac{Z_{c}}{\Omega}e^{-q^2 R_{c}^2 / 4-i\mathbf{q}\cdot\boldsymbol{\tau}^{\sigma}}$ and $\boldsymbol{\tau}^{\sigma}$ denotes the position of the $\sigma$ atom. $\Omega$ is the sample volume.

In the $(z, \mathbf{G})$ representation, the Fourier transform of the long-range core potential [the first term of Eq.~(\ref{eq:vlocdft})] is expressed as:
\begin{equation}\label{eq:FTlocLR}
V^{\text{LR}}_c(z, \mathbf{G}) = \frac{8\pi}{\Omega_c} \sum_{\sigma,\, g_z} \frac{Z_c^\sigma}{G^2 + g_z^2} \, e^{i g_z (z - \tau_z^\sigma)} \, e^{-\frac{(G^2 + g_z^2) R_c^2}{4} - i \mathbf{G} \cdot \boldsymbol{\tau}_\parallel^\sigma}
\end{equation}

Here, $\mathbf{G}$ denotes the in-plane two-dimensional reciprocal lattice vectors of the monolayer TMDC, and $g_z$ is the out-of-plane component of the reciprocal lattice vector for the supercell used. The volume of the supercell is given by $\Omega_c = A_c L_c$, where $A_c$ is the in-plane area and $L_c$ is the length along the out-of-plane direction of the supercell. $\boldsymbol{\tau}_\parallel^\sigma$ and $\tau_z^\sigma$ denote the in-plane and out-of-plane components of the position vector $\boldsymbol{\tau}^\sigma$ of atom $\sigma$ within the unit cell.

The second term in Eq.~(\ref{eq:vlocdft}) provides the short-range potential contribution. In the current SEP, we fit the $V_{c,SR}^{\sigma}(\mathbf{r})$ term for the metallic atom ($\sigma=M$) using the following expression:
\begin{equation}
V_{c,SR}^{M}(\mathbf{r}) =
\begin{cases}
    \sum\limits_{n=0}^{5} C_{n}^{M} r^{2n}   & r \leq r_1^M \\[10pt]
    \sum\limits_{n=0}^{5} C_{n}^{M} r^{2n} e^{-\alpha_M r^2},  & r^M_1 < r \leq r_2^M
\end{cases}
\label{eq:SRcore}
\end{equation}
and for the chalcogen atom ($\sigma=X$), we use:
\begin{equation}
V_{c,SR}^{X}(\mathbf{r}) =
\begin{cases}
    \sum\limits_{n=0}^{5} C_{n}^{X} r^{2n},  & r \leq r_1^X  \\[10pt]
    \sum\limits_{n=0}^{5} C_{n}^{X} r^{2n},  & r_1^X < r \leq r_2^{X} \\[10pt]
    \sum\limits_{n=0}^{5} C_{n}^{X} r^{2n},  & r_2^{X} < r \leq r_3^{X}
\end{cases}
\label{eq:SRcore1}
\end{equation}

\begin{figure}[h]
    \centering
    \includegraphics[width=\linewidth]{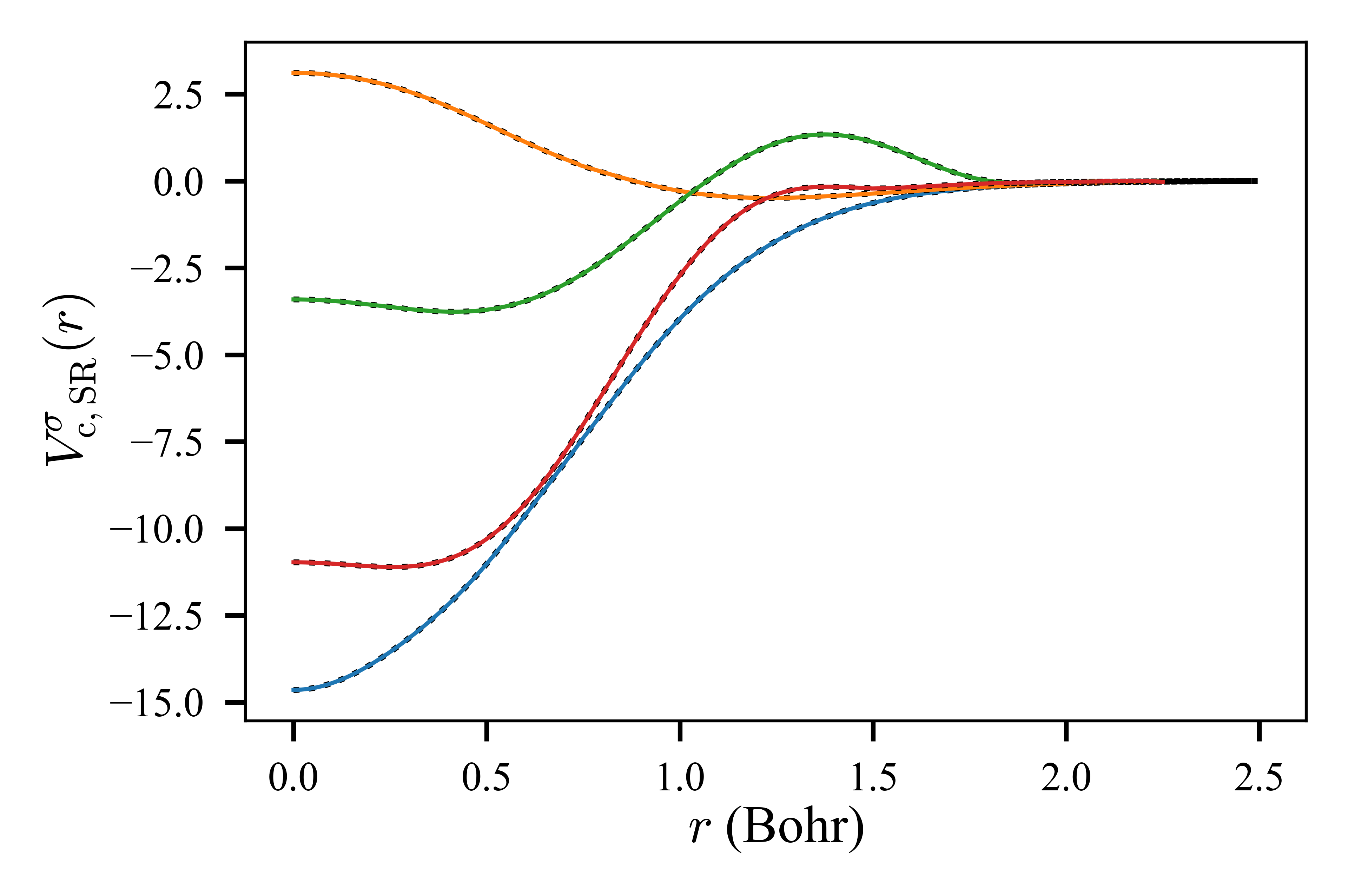}
    \caption{Comparison of the best-fit results for short-range interactions with those from the Vanderbilt ultrasoft pseudopotential \cite{vanderbilt1990soft} for W, Mo, S, and Se atoms. The solid-colored lines represent the fitted curves, while the black dots show the numerical results from the Vanderbilt ultrasoft pseudopotential (USPP).}
    \label{fig:vandSR}
\end{figure}

The best-fit results (solid lines) are shown in Fig.~\ref{fig:vandSR} together with the numerical results (dots) used in the Vanderbilt ultrasoft pseudopotential (USPP) \cite{vanderbilt1990soft}. The best-fit parameters are listed in Table~\ref{tab:tablelvSR}.

% ===================================================================
% TABLE — Short-range core potential parameters
% ===================================================================
\begin{table*}[h]
\centering
\footnotesize
\setlength{\tabcolsep}{4pt}
\renewcommand{\arraystretch}{1.0}
\caption{Best-fit parameters for the short-range core potentials of W, Mo, Se, and S atoms.
Dashes (--) indicate parameters not used.}
\label{tab:tablelvSR}
\begin{adjustbox}{max width=\linewidth}
\begin{tabular}{@{} c c
    D{.}{.}{1.5}
    D{.}{.}{1.5}
    D{.}{.}{3.5}
    D{.}{.}{3.5}
    D{.}{.}{3.5}
    D{.}{.}{3.5}
    D{.}{.}{3.5}
    D{.}{.}{3.5} @{}}
\toprule
\textbf{Atom} & \textbf{Zone}
  & \multicolumn{1}{c}{$r_{\text{cut}}$ (\AA)}
  & \multicolumn{1}{c}{$\alpha$}
  & \multicolumn{1}{c}{$C_0$}
  & \multicolumn{1}{c}{$C_1$}
  & \multicolumn{1}{c}{$C_2$}
  & \multicolumn{1}{c}{$C_3$}
  & \multicolumn{1}{c}{$C_4$}
  & \multicolumn{1}{c}{$C_5$} \\
\midrule
\multirow{2}{*}{W}
  & 1 & 0.74070 & \dash
    & 3.11326 & -5.84398 & -3.96957 & 20.39224 & -22.41975 & 8.54660 \\
  & 2 & 3.00000 & 1.92952
    & \dash & 10.49302 & -18.86608 & 8.95193 & -2.78125 & 0.23383 \\
\midrule
\multirow{2}{*}{Mo}
  & 1 & 0.74070 & \dash
    & -14.64721 & 20.11884 & -53.35016 & 194.88336 & -337.87171 & 212.53631 \\
  & 2 & 3.00000 & 2.65095
    & \dash & -93.30706 & 111.27523 & -109.04823 & 42.67719 & -7.80775 \\
\midrule
\multirow{3}{*}{Se}
  & 1 & 0.99864 & \dash
    & -3.40850 & -4.37431 & 15.07988 & -9.76815 & 1.57687 & 0.31197 \\
  & 2 & 1.88920 & \dash
    & -4.02380 & -1.14153 & 9.75924 & -6.71512 & 1.68661 & -0.14617 \\
  & 3 & 3.00000 & \dash
    & -3.02820 & 2.73354 & -1.03156 & 0.20143 & -0.02019 & 0.00083 \\
\midrule
\multirow{3}{*}{S}
  & 1 & 0.98996 & \dash
    & -10.97583 & -4.21172 & 34.92520 & -31.40929 & 8.56341 & 0.40383 \\
  & 2 & 1.49543 & \dash
    & -15.37907 & 14.59314 & 4.16391 & -9.25902 & 3.60809 & -0.44657 \\
  & 3 & 2.20320 & \dash
    & 11.36892 & -17.75451 & 10.48766 & -2.99507 & 0.41763 & -0.02286 \\
\bottomrule
\end{tabular}
\end{adjustbox}
\end{table*}

In the $(z, \mathbf{G})$ representation, we have
\begin{equation}
V^{SR}_{c}(z, \mathbf{G}) = \sum_{g_{z}} e^{i\mathbf{g}_{z} z} \widetilde{V}_{c,SR}^{\sigma}(|\mathbf{G}+\mathbf{g}_{z}|) e^{- i \mathbf{G} \cdot \boldsymbol{\tau}_\parallel^\sigma}
\label{eq:SRcore2_}
\end{equation}
with
\begin{equation}
\widetilde{V}_{c,SR}^{\sigma}(|\mathbf{G}+\mathbf{g}_{z}|)= \int r \, dr \, \sin\left( |\mathbf{G} + \mathbf{g}_{z}| r \right) \frac{V_{c,SR}^{\sigma}(r)}{|\mathbf{G}+ \mathbf{g}_{z}|}
\label{eq:SRcore2}
\end{equation}

Here, we have used the identity
\begin{equation}
e^{-i\mathbf{Q} \cdot \mathbf{r}} = 4\pi \sum_{l=0}^{\infty} \sum_{m=-l}^{l} (-i)^{l} j_{l}(Qr) Y_{lm}(\hat{\mathbf{Q}}) Y_{lm}^{\ast}(\hat{\mathbf{r}})
\end{equation}

Here, $\hat{\mathbf{Q}}$ represents the solid angle of the vector $\mathbf{Q} = \mathbf{G} + \mathbf{g}_{z}$ in spherical coordinates. $j_{l}(Qr)$ denotes the spherical Bessel function of order $l$. Since $V_{c,SR}^{\sigma}(r)$ has spherical symmetry, only the $l=0$ term survives. Using the analytic expressions in Eq.~(\ref{eq:SRcore}) and Eq.~(\ref{eq:SRcore1}), the integral in Eq.~(\ref{eq:SRcore2}) can be evaluated efficiently with the Gaussian quadrature method.

Combining Eqs.~\eqref{eq:FTlocLR} and~\eqref{eq:SRcore2_}, we write the ionic pseudopotential in the $(z, \mathbf{G})$ representation as
\begin{equation}
V^{\text{SEP}}_{\text{ion}}(z, \mathbf{G}) = V^{\text{LR}}_{c}(z, \mathbf{G}) + V^{\text{SR}}_{c}(z, \mathbf{G})
\label{eq:V_ion_SEP}
\end{equation}

\subsection{Hartree-exchange-correlation potential}
\label{sec:hxc_potential}
To determine \( V_{\text{hxc}}[\rho(\mathbf{r})] \), we begin by performing a self-consistent DFT calculation \cite{ren2015mixed}. This produces the ground-state electron density \( \rho(\mathbf{r}) \), from which we derive the effective local potential \( V_{\text{eff}}(\mathbf{r}) \). The hxc potential is then isolated by subtracting the core and atomic contributions. This step is essential because the hxc potential is specific to the system and is not transferable between different systems.

The potential \( V_{\text{hxc}}(z, \mathbf{G}) \) is derived by performing a Fourier transform of the real-space potential using the FFTW library \cite{frigo2005design}. Due to the point-group symmetry of the crystal, the reciprocal lattice vectors \( \mathbf{G} \) can be grouped into stars, where all vectors within a star are related by symmetry operations. Consequently, the hxc potential \( V_{\text{hxc}}(z, \mathbf{G}) \) for different \( \mathbf{G} \)'s in the same star can be related through the structure factor \( S(\mathbf{G}) \).

The structure factor of the 2D crystal is given by:
\begin{equation} \label{eq:strucfact}
S(\mathbf{G}) = S^{M}(\mathbf{G}) + S^{X}(\mathbf{G})
\end{equation}
where the contributions from the M (metal) and X (chalcogen) atoms are:
\begin{equation}
\left\{
\begin{aligned}
    S^{M}(\mathbf{G}) &= e^{i \mathbf{G} \cdot \boldsymbol{\tau}^{M}} = 1 \\
    S^{X}(\mathbf{G}) & = e^{i \mathbf{G} \cdot \boldsymbol{\tau}_{\parallel}}
\end{aligned}
\right.
\end{equation}
Here, we set the coordinate of the M atom at \( \boldsymbol{\tau}^{M} = (0,0,0) \), and the positions of the two X atoms are 
\( \boldsymbol{\tau}^{X_1} = \boldsymbol{\tau}_{\parallel} - \tau_c \hat{z} \) 
and 
\( \boldsymbol{\tau}^{X_2} = \boldsymbol{\tau}_{\parallel} + \tau_c \hat{z} \), 
which share the same in-plane coordinate 
\( \boldsymbol{\tau}_{\parallel} = (0, a/\sqrt{3}) \).

To model the hxc potential, we decompose it into long-range (LR) and short-range (SR) components:
\begin{equation}\label{eq:difVhxc}
V_{\text{hxc}}(z, \mathbf{G}) =  V_{\text{hxc}}^{\text{LR}}(z, \mathbf{G}) + V_{\text{hxc}}^{\text{SR}}(z, \mathbf{G})
\end{equation}
The first term denotes the long-range contributions, which are dominant for small \( \mathbf{G} \), while the second term corresponds to the remaining contribution, which is a slowly varying function of \( \mathbf{G} \) and thus short-ranged in real space.

\subsubsection{Short-range contribution} \label{sec:short_range}

The short-range component of the exchange-correlation potential, $V_{\text{hxc}}^{\text{SR}}(z, \mathbf{G})$, is determined from the self-consistent DFT calculations. Our analysis reveals two critical features across all four transition metal dichalcogenides (TMDCs): First, the imaginary component of $V_{\text{hxc}}^{\text{SR}}$ becomes negligible for reciprocal vectors beyond the fourth star ($|\mathbf{G}| > G_4$). Second, the $z$-dependence exhibits a Gaussian profile centered at the metal atomic plane ($z=0$), confirming the dominant contribution from the transition metal (M) atom.

This behavior motivates the following functional form:
\begin{equation} \label{eq:VHXC_SR_factored}
V_{\text{hxc}}^{\text{SR}}(z, \mathbf{G}) = D G^{4} e^{-b G^{2}} e^{-c z^{2}}
\end{equation}
where the prefactor $D G^{4} e^{-b G^{2}}$ governs the $\mathbf{G}$-dependence and $e^{-c z^{2}}$ describes the spatial decay perpendicular to the monolayer. The factor $G^4$ ensures minimal impact for small $\mathbf{G}$ wave vectors, while the decaying exponential $e^{-bG^2}$ maintains the softness required for efficient convergence. This formulation naturally vanishes at $\mathbf{G} = \mathbf{0}$, preserving the form of the leading term in the long-range potential.

\begin{table}[h]
\centering
\footnotesize
\caption{Universal short-range Hartree--exchange--correlation parameters for TMDC monolayers.}
\label{tab:tablesrhxc}
\begin{tabular}{c c c c}
\toprule
Material & $b$ & $c$ & $D$ \\
\midrule
MoS$_2$  & 0.14825 & 4.30923 & -0.00045 \\
MoSe$_2$ & 0.15644 & 4.68800 & -0.00050 \\
WS$_2$   & 0.21253 & 2.41437 & -0.00307 \\
WSe$_2$  & 0.21620 & 2.45912 & -0.00312 \\
\bottomrule
\end{tabular}
\end{table}

\begin{figure}[h]
\centering
\includegraphics[width=\columnwidth]{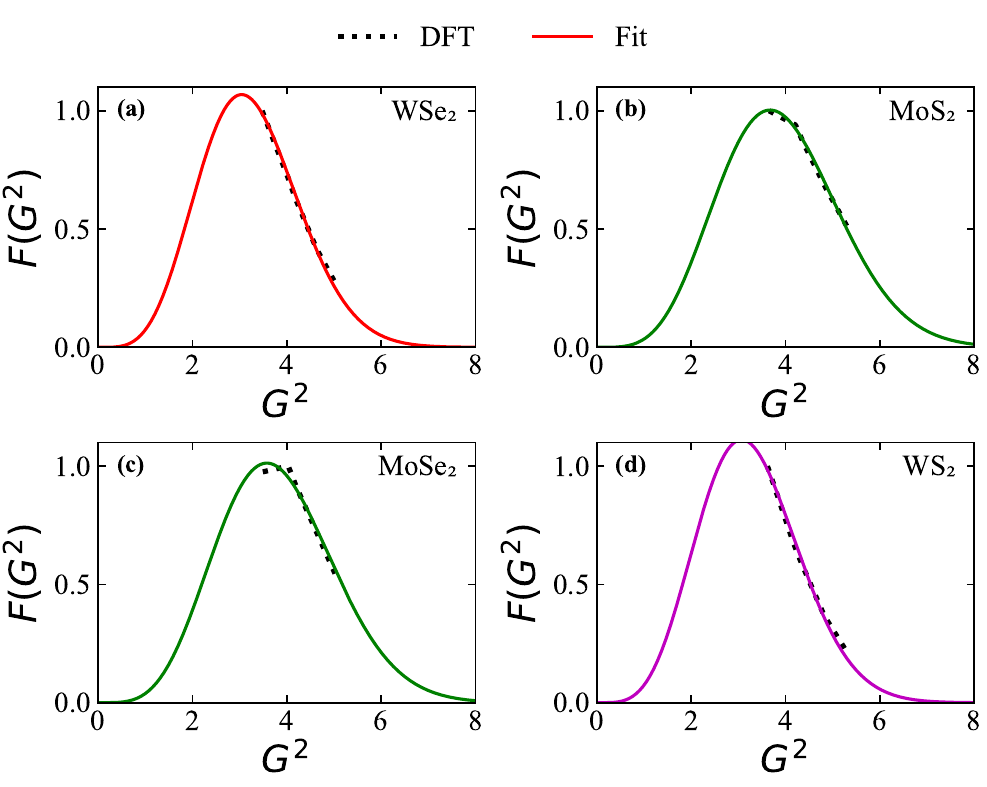}
\caption{(Color online) Scaling function $D G^{4} e^{-b G^{2}}$ for (a) WSe$_2$, (b) MoS$_2$, (c) MoSe$_2$, and (d) WS$_2$. Black dashed lines show the DFT results, while solid lines denote the fitted functions for each material. The data are averaged over stars of $\mathbf{G}$ with $s = 5, \ldots, 9$.}
\label{fig:VhxcSR_scaling}
\end{figure}

\begin{figure}[h]
\centering
\includegraphics[width=\columnwidth]{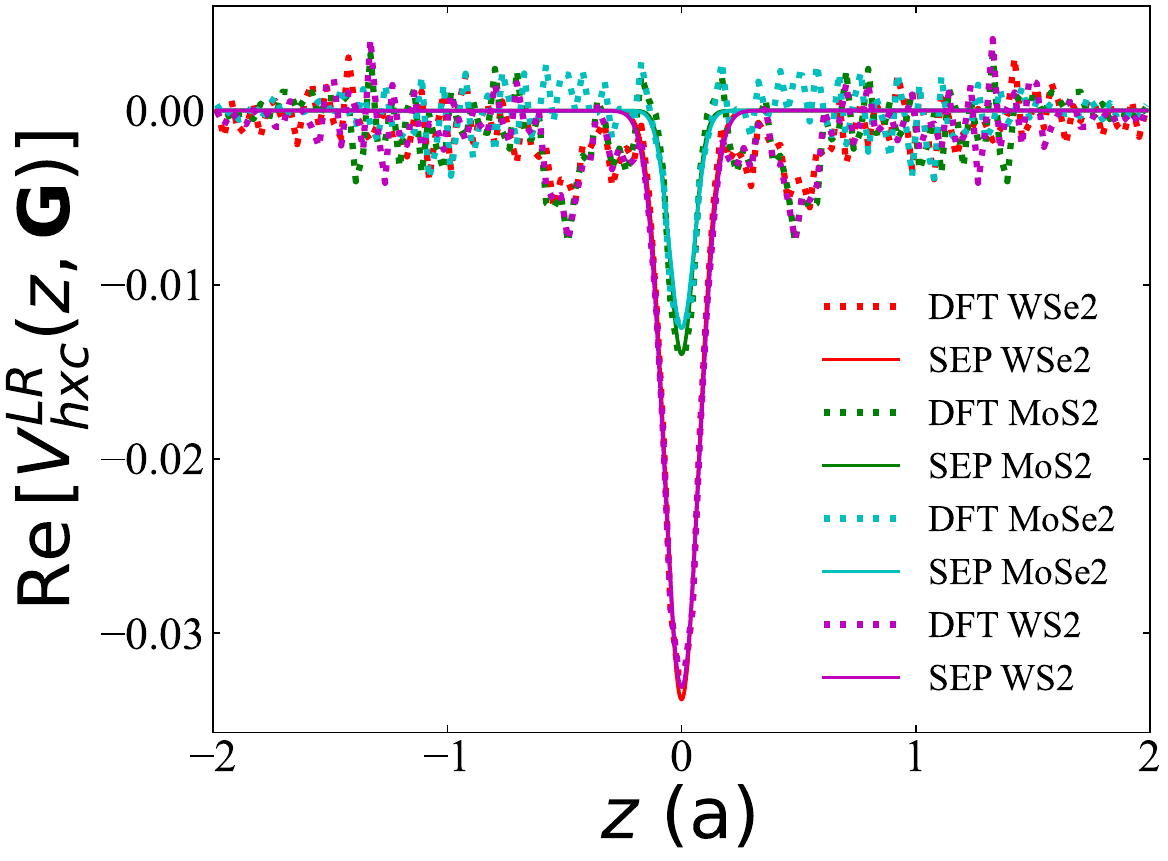}
\caption{(Color online) Universal $z$-dependence $e^{-c z^{2}}$ of the short-range potential. The $z$ coordinate is expressed in units of the lattice constant $a$. Dashed lines show the DFT results, while solid lines denote the fitted functions. The data are averaged over $\mathbf{G}$ vectors in stars 5--9.}
\label{fig:VhxcSR_univ}
\end{figure}

Fig.~\ref{fig:VhxcSR_scaling} shows the comparison between our scaling function and the $G$-dependence of the DFT-calculated potentials for $\mathbf{G}$-stars with index $s$ ranging from 5 to 9. The universal $z$-dependence is further validated in Fig.~\ref{fig:VhxcSR_univ}, where the Gaussian form $e^{-c z^{2}}$ excellently reproduces the DFT data. Table~\ref{tab:tablesrhxc} provides the best-fit parameters for all materials, showing systematic trends across the TMDC series.

\subsubsection{Long-range contribution}
For the first few \( \mathbf{G} \) stars (with index \( s \leq 4 \)), the long-range component is expressed as:
\begin{equation}
V^{\text{LR}}_{\text{hxc}}\left( z, \mathbf{G} \right)
= f^{M}(z)\, S^{M}\left( \mathbf{G} \right)
+ f^{X}(z)\, S^{X}\left( \mathbf{G} \right)
\end{equation}
where the shape functions \( f^{M}(z) \) and \( f^{X}(z) \) are given by sums of Gaussian functions:
\begin{equation}
\begin{aligned}
f^{M}(z) &= \sum_{i=1}^{2} A_{i}^{M} e^{-\alpha_{i}^{M} z^{2}}
+ B_{1}^{M} \left[
e^{-\beta_{1}^{M} ( z - \tau_c )^{2}}
+ e^{-\beta_{1}^{M} ( z + \tau_c )^{2}}
\right] \\
f^{X}(z) &= \sum_{i=1}^{3} A_{i}^{X} e^{-\alpha_{i}^{X} z^{2}}
+ \sum_{i=1}^{2} B_{i}^{X} \left[
e^{-\beta_{i}^{X} ( z- \tau_c )^{2}}
+ e^{-\beta_{i}^{X} ( z + \tau_c)^{2}}
\right]
\end{aligned}
\end{equation}

Here, \( f^{M}(z) \) and \( f^{X}(z) \) represent the contributions from the M and X atomic species, respectively. The parameters \( A^{M}_{i}(A^{X}_{i}) \), \( B^{M}_{1}(B^{X}_{i} ) \), \( \alpha^{M}_{i} (\alpha^{X}_{i})\), \( \beta^{M}_{1} \), and \( \beta^{X}_{i} \) are determined by fitting to the DFT-calculated HXC potential after subtracting the short-range contribution in Eq.~\eqref{eq:VHXC_SR_factored}. The best-fit parameters for the TMDCs are presented in Tables~\ref{tab:alpha_params} and~\ref{tab:A_params}.

% ===================================================================
% TABLE 3a — Shape parameters (alpha / beta)
% ===================================================================
\begin{table*}[ht]
\centering
\caption{Parameters $\boldsymbol\alpha_i^{\sigma}$ and $\boldsymbol\beta_i^{\sigma}$ used in shape functions
$f^{\sigma}(\mathbf{G})$ for fitting the long-range exchange-correlation potential for
WSe$_2$, MoSe$_2$, MoS$_2$, and WS$_2$.
The superscript $\sigma$ denotes the atomic species.
Dashes (--) indicate parameters that are either zero or not used.}
\label{tab:alpha_params}
\footnotesize
\setlength{\tabcolsep}{6.5pt}
\renewcommand{\arraystretch}{1.0}
\begin{tabular}{@{} l c c r r r r r @{}}
\toprule
\textbf{Material} & \textbf{G-star} ($G_s$) & \textbf{Atom}
  & \multicolumn{1}{c}{$\alpha^\sigma_1$}
  & \multicolumn{1}{c}{$\alpha^\sigma_2$}
  & \multicolumn{1}{c}{$\alpha^\sigma_3$}
  & \multicolumn{1}{c}{$\beta^\sigma_1$}
  & \multicolumn{1}{c}{$\beta^\sigma_2$} \\
\midrule

%--- WSe2
\multirow{10}{*}{WSe$_2$}
  & \multirow{2}{*}{$0$} & W
    & 0.05270 & 0.03990 & 0.05270 & 0.08800 & \dash \\
  & & Se
    & \dash & \dash & \dash & 0.05240 & 0.62640 \\
\cmidrule{2-8}
  & \multirow{2}{*}{$G_1$} & W
    & 3.63780 & 0.29220 & \dash & 0.33930 & \dash \\
  & & Se
    & 0.28570 & 0.26280 & 0.30880 & 0.24380 & 0.98210 \\
\cmidrule{2-8}
  & \multirow{2}{*}{$G_2$} & W
    & 0.84020 & 0.83200 & \dash & 2.59800 & \dash \\
  & & Se
    & 0.03780 & \dash & \dash & 0.14720 & 0.32590 \\
\cmidrule{2-8}
  & \multirow{2}{*}{$G_3$} & W
    & 2.25500 & 0.37620 & \dash & 0.63970 & \dash \\
  & & Se
    & 0.23000 & \dash & \dash & 0.69190 & 0.12110 \\
\cmidrule{2-8}
  & \multirow{2}{*}{$G_4$} & W
    & 2.24970 & 0.14520 & \dash & 0.78200 & \dash \\
  & & Se
    & 0.43000 & \dash & \dash & 2.21670 & 0.05060 \\
\midrule

%--- MoSe2
\multirow{10}{*}{MoSe$_2$}
  & \multirow{2}{*}{$0$} & Mo
    & 0.00990 & 0.01000 & 0.11170 & 0.08650 & \dash \\
  & & Se
    & \dash & \dash & \dash & 0.01810 & 0.27590 \\
\cmidrule{2-8}
  & \multirow{2}{*}{$G_1$} & Mo
    & 20.14220 & 0.30180 & \dash & 0.39770 & \dash \\
  & & Se
    & 0.29090 & 0.26610 & 0.31440 & 0.24830 & 0.96060 \\
\cmidrule{2-8}
  & \multirow{2}{*}{$G_2$} & Mo
    & 0.78590 & 0.77570 & \dash & 2.66080 & \dash \\
  & & Se
    & 0.03990 & \dash & \dash & 0.21570 & 0.22070 \\
\cmidrule{2-8}
  & \multirow{2}{*}{$G_3$} & Mo
    & 0.42990 & 5.85410 & \dash & 0.76240 & \dash \\
  & & Se
    & 0.03940 & \dash & \dash & 0.08980 & 0.86300 \\
\cmidrule{2-8}
  & \multirow{2}{*}{$G_4$} & Mo
    & 2.11930 & 1.96880 & \dash & 0.39970 & \dash \\
  & & Se
    & 0.44700 & \dash & \dash & 0.04900 & 2.29270 \\
\midrule

%--- MoS2
\multirow{10}{*}{MoS$_2$}
  & \multirow{2}{*}{$0$} & Mo
    & 0.05900 & 0.04380 & 1.35560 & 0.09530 & \dash \\
  & & S
    & \dash & \dash & \dash & 0.05760 & 0.72540 \\
\cmidrule{2-8}
  & \multirow{2}{*}{$G_1$} & Mo
    & 13.27200 & 0.31700 & \dash & 0.39420 & \dash \\
  & & S
    & 0.32270 & 0.30760 & 0.33980 & 0.26270 & 3.00030 \\
\cmidrule{2-8}
  & \multirow{2}{*}{$G_2$} & Mo
    & 0.54750 & 2.83810 & \dash & 0.09660 & \dash \\
  & & S
    & 0.12550 & \dash & \dash & 1.07170 & 1.09020 \\
\cmidrule{2-8}
  & \multirow{2}{*}{$G_3$} & Mo
    & 5.80840 & 0.42340 & \dash & 0.73030 & \dash \\
  & & S
    & 0.03820 & \dash & \dash & 0.08570 & 2.34220 \\
\cmidrule{2-8}
  & \multirow{2}{*}{$G_4$} & Mo
    & 2.14070 & 2.05410 & \dash & 0.57270 & \dash \\
  & & S
    & 0.02170 & \dash & \dash & 0.03670 & 4.35380 \\
\midrule

%--- WS2
\multirow{10}{*}{WS$_2$}
  & \multirow{2}{*}{$0$} & W
    & 0.09070 & 0.05496 & 0.11409 & 0.10271 & \dash \\
  & & S
    & \dash & \dash & \dash & 0.07126 & 0.16743 \\
\cmidrule{2-8}
  & \multirow{2}{*}{$G_1$} & W
    & 3.11381 & 0.31103 & \dash & 0.33436 & \dash \\
  & & S
    & 0.35731 & 0.35527 & 0.36652 & 0.25769 & 3.18066 \\
\cmidrule{2-8}
  & \multirow{2}{*}{$G_2$} & W
    & 0.54746 & 2.83811 & \dash & 0.09656 & \dash \\
  & & S
    & 0.02668 & \dash & \dash & 3.45239 & 0.12636 \\
\cmidrule{2-8}
  & \multirow{2}{*}{$G_3$} & W
    & 2.10427 & 0.40701 & \dash & 0.62162 & \dash \\
  & & S
    & 0.03292 & \dash & \dash & 2.10817 & 0.07401 \\
\cmidrule{2-8}
  & \multirow{2}{*}{$G_4$} & W
    & 2.12381 & 0.22227 & \dash & 0.09139 & \dash \\
  & & S
    & 1.37981 & \dash & \dash & 2.87953 & 2.87953 \\
\bottomrule
\end{tabular}
\end{table*}

% ===================================================================
% TABLE 3b — Amplitude parameters (A, in Ry)
% ===================================================================
\begin{table*}[ht]
\centering
\caption{Amplitude parameters $\mathbf{A}_i^\sigma$ and $\mathbf{B}_i^\sigma$ (in Ry) of the Gaussian
functions used in shape functions $f^\sigma(\mathbf{G})$ for fitting the long-range
exchange-correlation potential for WSe$_2$, MoSe$_2$, MoS$_2$, and WS$_2$.
Superscripts $\sigma$ denote the atomic species.
Dashes (--) indicate parameters that are either zero or not applicable.}
\label{tab:A_params}
\footnotesize
\setlength{\tabcolsep}{6.0pt}
\renewcommand{\arraystretch}{1.0}
\begin{tabular}{@{} l c c r r r r r @{}}
\toprule
\textbf{Material} & \textbf{G-star} ($G_s$) & \textbf{Atom}
  & \multicolumn{1}{c}{$A^\sigma_1$}
  & \multicolumn{1}{c}{$A^\sigma_2$}
  & \multicolumn{1}{c}{$A^\sigma_3$}
  & \multicolumn{1}{c}{$B^\sigma_1$}
  & \multicolumn{1}{c}{$B^\sigma_2$} \\
\midrule

%--- WSe2
\multirow{10}{*}{WSe$_2$}
  & \multirow{2}{*}{$0$} & W
    & 5927.2 & -7545.6 & 69.486 & 158.843 & \dash \\
  & & Se
    & \dash & \dash & \dash & 1239.5 & 0.16300 \\
\cmidrule{2-8}
  & \multirow{2}{*}{$G_1$} & W
    & -0.05623 & 0.90370 & \dash & 0.04182 & \dash \\
  & & Se
    & 5.95130 & -2.76040 & -3.13980 & 0.16790 & -0.06851 \\
\cmidrule{2-8}
  & \multirow{2}{*}{$G_2$} & W
    & -13.36220 & 13.43910 & \dash & -0.02067 & \dash \\
  & & Se
    & -0.01691 & \dash & \dash & 0.04048 & -0.03000 \\
\cmidrule{2-8}
  & \multirow{2}{*}{$G_3$} & W
    & -0.03250 & 0.05040 & \dash & -0.00524 & \dash \\
  & & Se
    & -0.00984 & \dash & \dash & -0.03036 & 0.01137 \\
\cmidrule{2-8}
  & \multirow{2}{*}{$G_4$} & W
    & -0.00545 & 0.00270 & \dash & 0.00399 & \dash \\
  & & Se
    & 0.00399 & \dash & \dash & -0.00552 & -0.00243 \\
\midrule

%--- MoSe2
\multirow{10}{*}{MoSe$_2$}
  & \multirow{2}{*}{$0$} & Mo
    & 36915.0 & -37872.3 & 33.5467 & 18.9963 & \dash \\
  & & Se
    & \dash & \dash & \dash & 567.569 & 1.42313 \\
\cmidrule{2-8}
  & \multirow{2}{*}{$G_1$} & Mo
    & -0.01117 & 0.96784 & \dash & 0.04602 & \dash \\
  & & Se
    & 4.19276 & -1.86744 & -2.27322 & 0.16809 & -0.07188 \\
\cmidrule{2-8}
  & \multirow{2}{*}{$G_2$} & Mo
    & -7.45334 & 7.56968 & \dash & -0.02077 & \dash \\
  & & Se
    & -0.01551 & \dash & \dash & 1.13877 & -1.12930 \\
\cmidrule{2-8}
  & \multirow{2}{*}{$G_3$} & Mo
    & 0.05440 & -0.01320 & \dash & -0.00389 & \dash \\
  & & Se
    & -0.01548 & \dash & \dash & 0.01590 & -0.02536 \\
\cmidrule{2-8}
  & \multirow{2}{*}{$G_4$} & Mo
    & -0.23349 & 0.23095 & \dash & 0.00084 & \dash \\
  & & Se
    & 0.00423 & \dash & \dash & -0.00231 & -0.00539 \\
\midrule

%--- MoS2
\multirow{10}{*}{MoS$_2$}
  & \multirow{2}{*}{$0$} & Mo
    & 5906.6 & -7597.70 & 0.08853 & 186.204 & \dash \\
  & & S
    & \dash & \dash & \dash & 1321.1 & 0.23282 \\
\cmidrule{2-8}
  & \multirow{2}{*}{$G_1$} & Mo
    & -0.01469 & 0.93427 & \dash & 0.04288 & \dash \\
  & & S
    & 3.87226 & -1.63097 & -2.18468 & 0.18556 & -0.01713 \\
\cmidrule{2-8}
  & \multirow{2}{*}{$G_2$} & Mo
    & 0.12022 & -0.03303 & \dash & 0.01033 & \dash \\
  & & S
    & -0.00156 & \dash & \dash & 1.12476 & -1.13208 \\
\cmidrule{2-8}
  & \multirow{2}{*}{$G_3$} & Mo
    & -0.01347 & 0.04308 & \dash & -0.00419 & \dash \\
  & & S
    & -0.01834 & \dash & \dash & 0.01719 & -0.01181 \\
\cmidrule{2-8}
  & \multirow{2}{*}{$G_4$} & Mo
    & -0.32737 & 0.32449 & \dash & 0.00073 & \dash \\
  & & S
    & 0.02460 & \dash & \dash & -0.01671 & -0.00664 \\
\midrule

%--- WS2
\multirow{10}{*}{WS$_2$}
  & \multirow{2}{*}{$0$} & W
    & 76299.8 & -70839.1 & -20287.9 & 7719.83 & \dash \\
  & & S
    & \dash & \dash & \dash & 7979.81 & 234.606 \\
\cmidrule{2-8}
  & \multirow{2}{*}{$G_1$} & W
    & -0.07260 & 0.88216 & \dash & 0.04101 & \dash \\
  & & S
    & 3.86050 & -1.63159 & -2.17128 & 0.18788 & -0.01614 \\
\cmidrule{2-8}
  & \multirow{2}{*}{$G_2$} & W
    & 0.12022 & -0.03303 & \dash & 0.01033 & \dash \\
  & & S
    & -0.00913 & \dash & \dash & -0.01313 & 0.01190 \\
\cmidrule{2-8}
  & \multirow{2}{*}{$G_3$} & W
    & -0.05057 & 0.03651 & \dash & -0.00429 & \dash \\
  & & S
    & -0.02078 & \dash & \dash & -0.00984 & 0.01798 \\
\cmidrule{2-8}
  & \multirow{2}{*}{$G_4$} & W
    & -0.00655 & 0.00191 & \dash & 0.00049 & \dash \\
  & & S
    & 0.00077 & \dash & \dash & 0.58232 & -0.58972 \\
\bottomrule
\end{tabular}
\end{table*}

After performing the SEP fitting to the DFT data, we observe that the potentials for the first two G-vector stars show slight deviations across all materials (MoS$_2$, WS$_2$, MoSe$_2$, and WSe$_2$). These deviations, though small, are crucial for achieving accurate band-structure calculations. To capture these differences, we define the residual potential as  
\[
\Delta V^{\text{LR}}_{\text{hxc}}\left( z, \mathbf{G} \right) = V_{\text{hxc}}^{\text{DFT}}\left( z, \mathbf{G} \right) - V_{\text{hxc}}^{\text{SR}}\left( z, \mathbf{G} \right) - V_{\text{hxc}}^{\text{LR}}\left( z, \mathbf{G} \right)
\]
which represents the discrepancy between the small-$\mathbf{G}$ components. To account for this residual, we fit the difference using the following functional forms:
\begin{equation}
\begin{aligned}
\Delta f_{\text{LR}}^{M}(z) &= p^{M} e^{-\alpha^{M} z^{2}} \cos\left(Qz\right) \\
\Delta f_{\text{LR}}^{X}(z) &= p^{X} \left[ e^{-\alpha^{X} \left( z - \tau_{c} \right)^{2}} + e^{-\alpha^{X} \left( z + \tau_{c} \right)^{2}} \right]
\end{aligned}
\end{equation}
where \(p^{M} (p^{X})\), \(\alpha^{M}(\alpha^{X})\) and \(Q\) are the fitting parameters. The parameters \(p^{M}\) and \(p^{X}\) are scaling factors, \(\alpha^{M}\) and \(\alpha^{X}\) control the width of the Gaussian decay, and \(Q\) captures any oscillatory behavior. These functions effectively model the observed deviations, enabling a more accurate representation of the potential. The best-fit parameters for each material are summarized in Table~\ref{tab:table4}.

\begin{figure*}[!t]
\centering
\includegraphics[width=\textwidth]{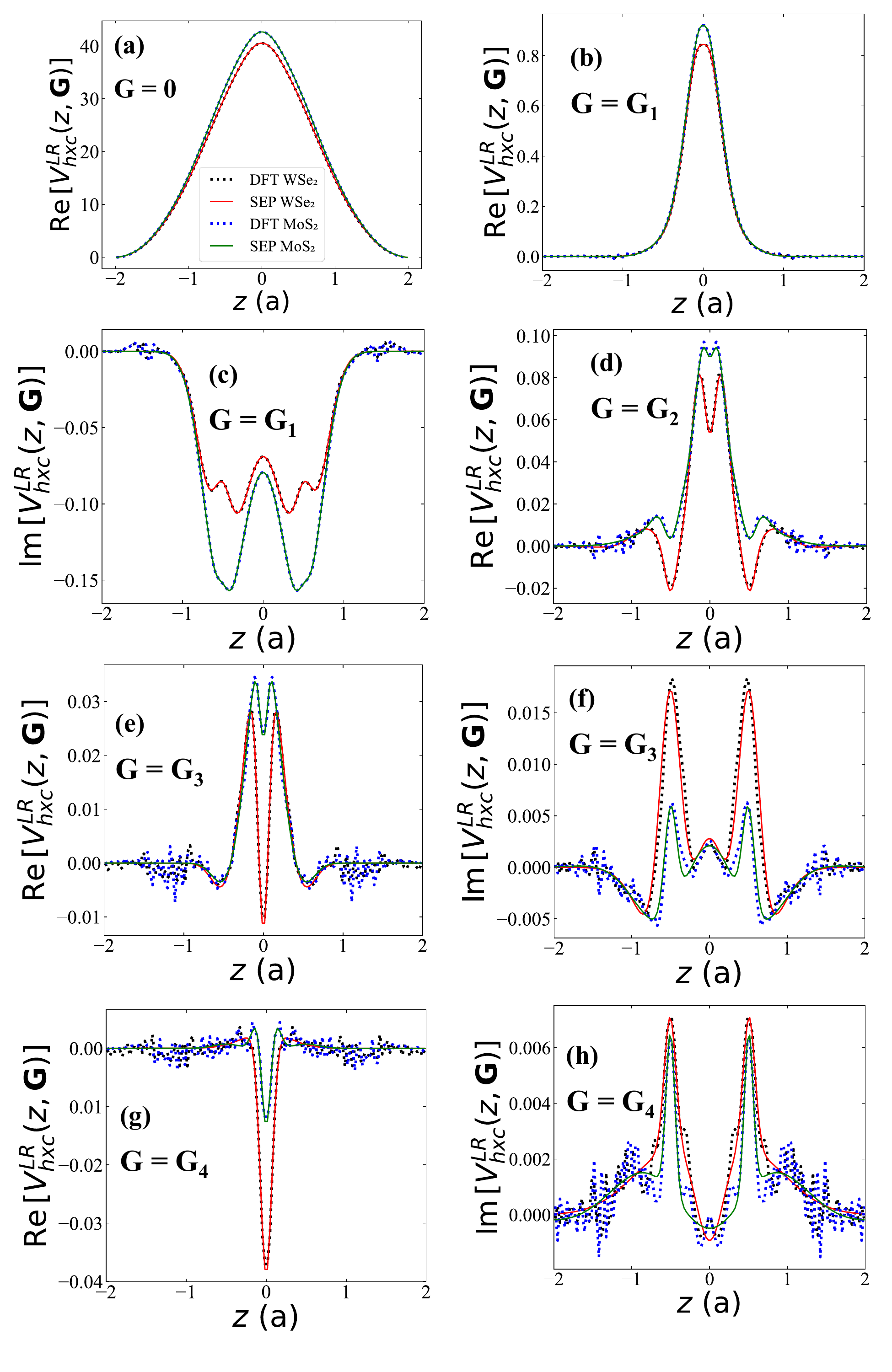}
\caption{Fourier transform of the effective local pseudopotential for reciprocal lattice vectors \( \mathbf{G} \) in the first few stars for WSe\textsubscript{2} and MoS\textsubscript{2}. The magnitudes of \( \mathbf{G} \) (in units of $2\pi/a$) are: (a) \( G = 0 \), (b,c) \( G = G_1 = 2/\sqrt{3} \), (d) \( G = G_2 = 2 \), (e,f) \( G = G_3 = 4/\sqrt{3} \), and (g,h) \( G = G_4 = 2\sqrt{7}/\sqrt{3} \), corresponding to stars with index $s = 0, 1, 2, 3, 4$. Fitted results (solid lines) are compared against DFT input potentials (dotted curves).  z coordinate (horigental) is in units of the lattice constant a.}
\label{fig:LRfit_WSe2_MoS2}
\end{figure*}

\begin{figure*}[!t]
\centering
\includegraphics[width=\textwidth]{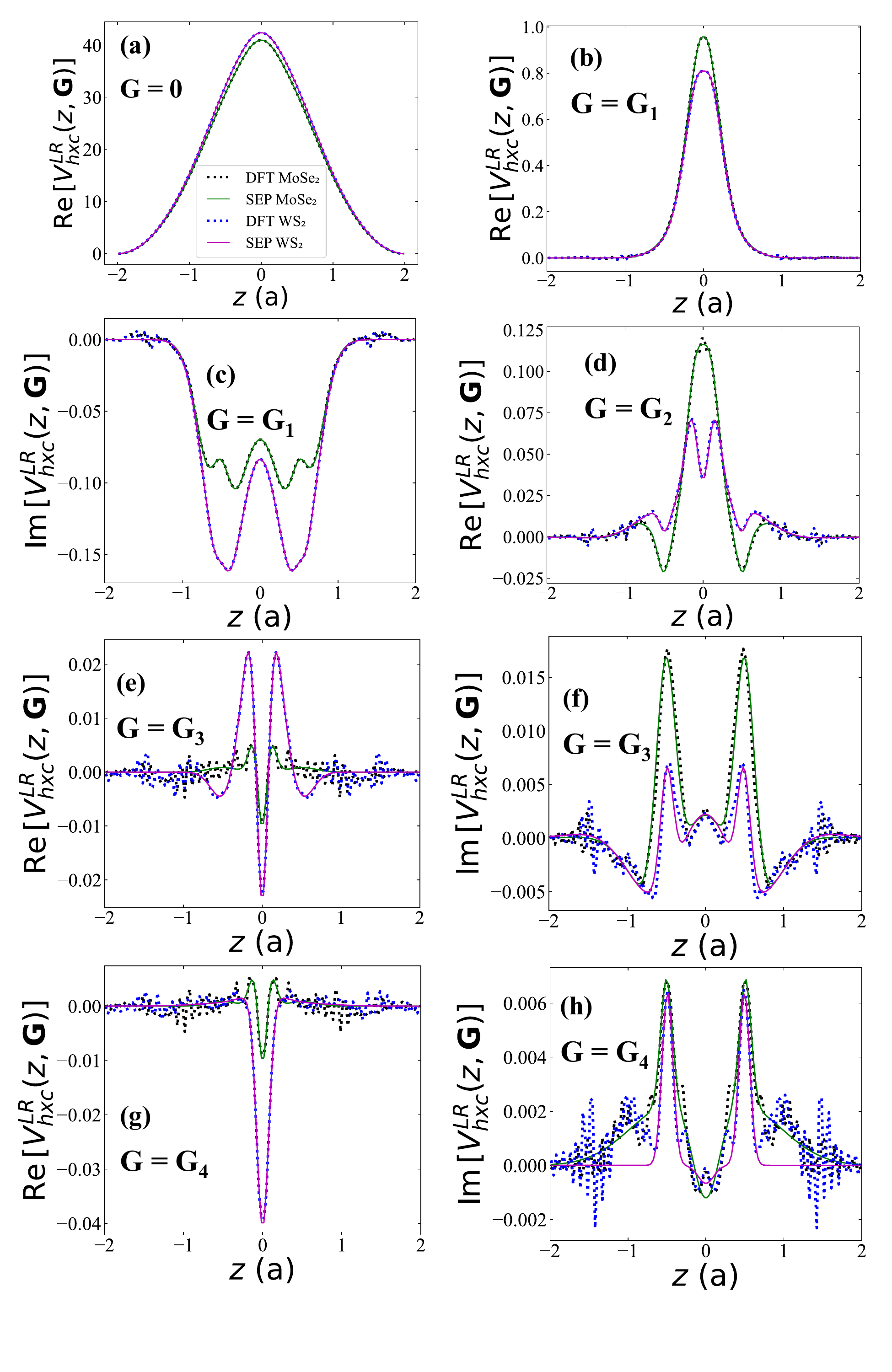}
\caption{Fourier transform of the effective local pseudopotential for reciprocal lattice vectors \( \mathbf{G} \) in the first few stars for MoSe\textsubscript{2} and WS\textsubscript{2}. The magnitudes of \( \mathbf{G} \) (in units of $2\pi/a$) are: (a) \( G = 0 \), (b,c) \( G = G_1 = 2/\sqrt{3} \), (d) \( G = G_2 = 2.0 \), (e,f) \( G = G_3 = 4/\sqrt{3} \), and (g,h) \( G = G_4 = 2\sqrt{7}/\sqrt{3} \), corresponding to stars with index $s = 0, 1, 2, 3, 4$. Fitted results (solid lines) are compared against DFT input potentials (dotted curves). z coordinate (horigental) is in units of the lattice constant a.}
\label{fig:LRfit_MoSe2_WS2}
\end{figure*}

Our net hxc contribution to the SEP is defined as
\begin{equation}
V^{\text{SEP}}_{\text{hxc}}\left( z, \mathbf{G} \right) =  V_{\text{hxc}}^{\text{SR}}\left( z, \mathbf{G} \right) + V_{\text{hxc}}^{\text{LR}}\left( z, \mathbf{G} \right) + \Delta V^{\text{LR}}_{\text{hxc}}\left( z, \mathbf{G} \right)
\label{eq:V_hxc_SEP}
\end{equation}
which agrees well with the potential obtained from the self-consistent DFT calculation \cite{ren2015mixed}. A comparison between our hxc potential, $V_{\text{hxc}}^{\text{SEP}}(z, \mathbf{G})$, and the DFT-calculated potential, $V_{\text{hxc}}^{\text{DFT}}(z, \mathbf{G})$, for the first few $\mathbf{G}$-vector stars is shown in Fig.~\ref{fig:LRfit_WSe2_MoS2} and Fig.~\ref{fig:LRfit_MoSe2_WS2}.

\begin{table}[h]
\centering
\footnotesize
\caption{Residual long-range Hartree--exchange--correlation parameters
$\Delta V^{\mathrm{LR}}_{\mathrm{hxc}}$ extracted from the SEP fitting.}
\label{tab:table4}
\begin{tabular}{@{} c c
    D{.}{.}{2.5}
    D{.}{.}{2.5}
    D{.}{.}{2.5}
    D{.}{.}{2.5}
    D{.}{.}{2.5} @{}}
\toprule
\textbf{Material} & \textbf{$G_s$}
  & \multicolumn{1}{c}{$p^{M}$}
  & \multicolumn{1}{c}{$\alpha^{M}$}
  & \multicolumn{1}{c}{$Q$}
  & \multicolumn{1}{c}{$p^{X}$}
  & \multicolumn{1}{c}{$\alpha^{X}$} \\
\midrule
WSe$_2$
  & $0$   & 0.00983 & 0.06514 & 2.40458 & -0.06000 & 3.31118 \\
\midrule
\multirow{2}{*}{MoSe$_2$}
  & $0$   & 0.02665 & 0.09183 & 2.44486 & -0.01356 & 0.16895 \\
  & $G_1$ & 0.00915 & 0.08428 & 2.46345 & -0.01354 & 1.70000 \\
\midrule
\multirow{2}{*}{MoS$_2$}
  & $0$   & 0.01696 & 0.09000 & 2.17832 & -0.07075 & 4.35178 \\
  & $G_1$ & 0.01324 & 0.56977 & 2.06550 & \dash & \dash \\
\midrule
\multirow{2}{*}{WS$_2$}
  & $0$   & -0.01502 & 0.19884 & 3.86680 & -0.06495 & 3.53613 \\
  & $G_1$ & 0.00301  & 0.03639 & 1.03575 & \dash & \dash \\
\bottomrule
\end{tabular}
\end{table}

Combining \eqref{eq:V_ion_SEP} and \eqref{eq:V_hxc_SEP}, we obtain the net semi-empirical local pseudopotential in the $(z, \mathbf{G})$ representation as
\begin{equation}\label{eq:localpot}
V^{\text{SEP}}_{\text{loc}}(z, \mathbf{G}) = V^{\text{SEP}}_{\text{ion}}(z, \mathbf{G}) + V^{\text{SEP}}_{\text{hxc}}(z, \mathbf{G})
\end{equation}

\subsection{Nonlocal pseudopotential}

Nonlocal pseudopotentials (NLPPs) are essential for accurately modeling the electronic structures of layered TMDCs, which involve transition metals. They capture the angular-momentum-dependent electron-ion interactions crucial for d-orbitals. Including NLPPs improves the description of chemical bonding, hybridization, and the overall electronic structure, enabling better predictions of bandgaps compared to using only local pseudopotentials.

In the scalar relativistic approximation, where spin–orbit coupling is neglected, the separable formulation introduced by Kleinman and Bylander \cite{kleinman1982efficacious} (KB) is given by
\begin{equation}\label{eq:nonlocalpot}
\widehat{V}_{nl} = \sum_{\sigma lm, nn'} E_{lm}^{nn'} \left| \beta_{lm}^{n\sigma} \right\rangle \left\langle \beta_{lm}^{n'\sigma} \right|
\end{equation}
Here, the projector functions \(\beta_{lm}^{n\sigma}\) and the coefficients \(E_{lm}^{nn'}\) characterize the non-local pseudopotentials for the various atomic species (Mo, W, S, Se). The \(\beta\) functions take the form:
\begin{equation}
\beta_{lm}^{n\sigma} \left( \mathbf{r} \right) = A_{l}^{n\sigma}(r) r^{l} Y_{lm}(\hat{\mathbf{r}})
\end{equation}
We fit the radial part $A_{l}^{n\sigma}(r)$ of the beta functions for each orbital, benchmarked against data from Refs. \cite{vanderbilt1990soft,garrity2014pseudopotentials}. For efficient fitting, we decompose the fitting function into two parts, defined in intervals $(0, R_s)$ and $(R_s, R_{\text{cut}})$, respectively:
\begin{equation}\label{eq:combined}
A_{l}^{n\sigma}(r) =
\begin{cases} 
\displaystyle \sum_{p=0}^{5} B^{\sigma}_{p}\, r^{2p} e^{-\alpha_{a}^{\sigma} r^{2}}, & \text{for } r < R_{s} \\[4mm]
\displaystyle \sum_{q=0}^{3} C^{\sigma}_{q}\, r^{2q} e^{-\alpha_{b}^{\sigma} r^{2}}, & \text{for } r \geq R_{s}
\end{cases}
\end{equation}

\begin{figure*}[t!] 
\centering
\includegraphics[width=\textwidth]{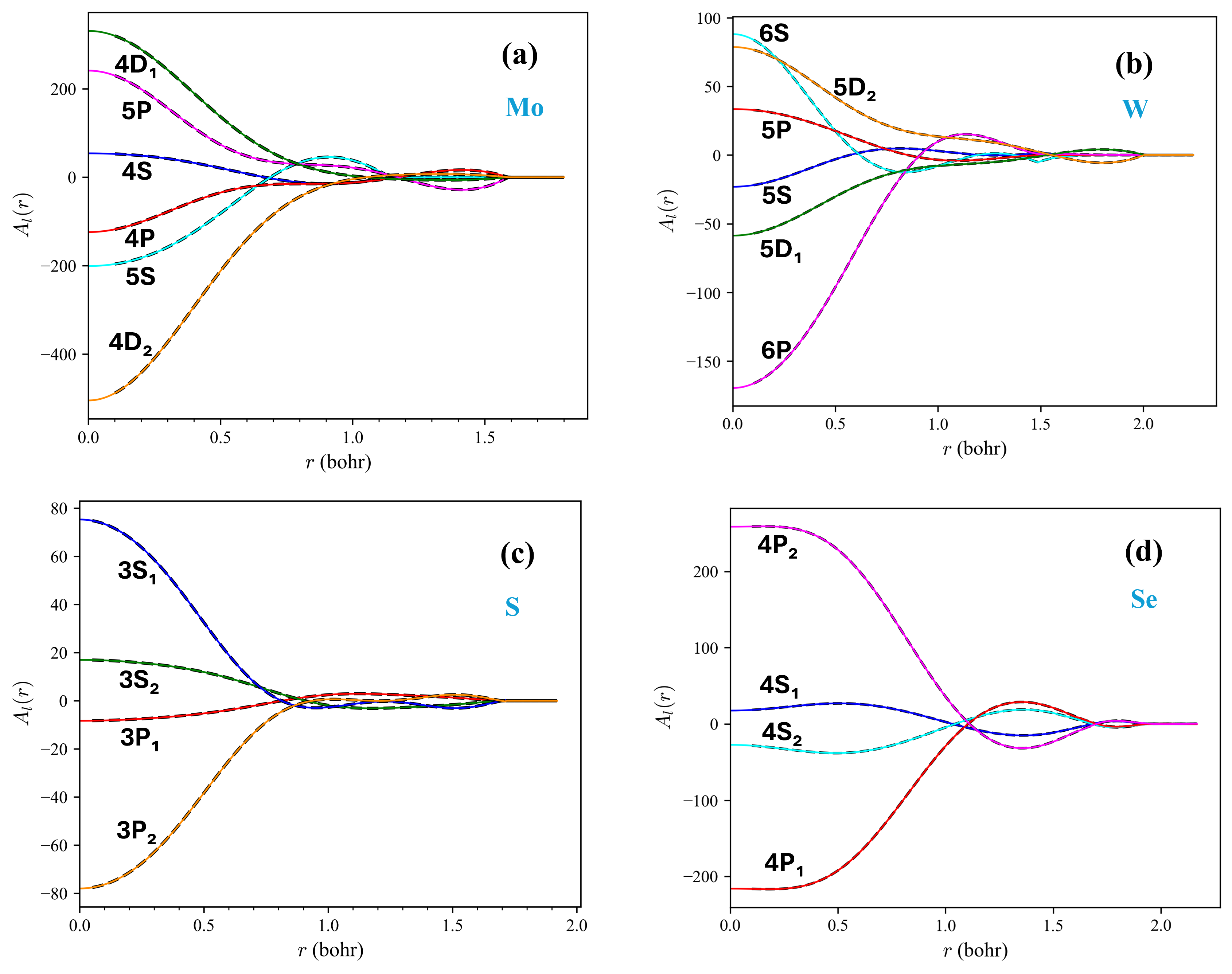}
\caption{Comparisons of best-fit $A_{l}^{n\sigma}(r)$ functions in the non-local pseudopotentials for (a) Mo, (b) W, (c) S, and (d) Se atoms. Colored lines represent the Semi-empirical Pseudopotential (SEP) model fits for individual orbitals, benchmarked against the black dashed lines showing corresponding results from Ultrasoft Pseudopotentials (USPP) \cite{vanderbilt1990soft,garrity2014pseudopotentials} used in the DFT calculation.}
\label{fig:betafit_plot}
\end{figure*}

Here, \( B^{\sigma}_{p} \), \( C^{\sigma}_q \), \( \alpha_a^{\sigma} \), % <-- verify: was \alpha
and \( \alpha_b^{\sigma} \) are fitting parameters.
The best-fit parameters for atoms M (Mo, W) and X (S, Se) are
provided in Tables~\ref{tab:beta_B} (for $r<R_s$) and~\ref{tab:beta_C} (for $r \geq R_s$).
These parameters were optimized via variational minimization,
resulting in a root-mean-square (RMS) fitting error below 1\%
across all orbital channels.
The comparison of our best-fit \( A_{l}^{n\sigma}(r) \) functions
with the reference data is shown in Fig.~\ref{fig:betafit_plot}
for (a) Mo, (b) W, (c) S, and (d) Se atoms.

% ===================================================================
% TABLE (a) — alpha + rcut1 + B amplitudes
% ===================================================================
\begin{table*}[t]
\centering
\footnotesize
\setlength{\tabcolsep}{4pt}
\renewcommand{\arraystretch}{1.0}
\caption{Best-fit parameters for the nonlocal pseudopotential beta functions of
Mo, W, S, and Se defined in  Eq. (\ref{eq:combined}) for $r<R_s$.  $\alpha_a^{\sigma}$, inner cutoff
$r_{\rm cut1}^\sigma$, and Gaussian amplitude parameters $B^\sigma$.
Dashes (--) indicate zero or unused.
All values rounded to five decimal places.}
\label{tab:beta_B}
\begin{adjustbox}{max width=\linewidth}
\begin{tabular}{@{} c c
    D{.}{.}{1.5} D{.}{.}{1.5}
    D{.}{.}{3.5} D{.}{.}{3.5} D{.}{.}{3.5}
    D{.}{.}{3.5} D{.}{.}{3.5} D{.}{.}{3.5} @{}}
\toprule
\textbf{Atom} & $l$
  & \multicolumn{1}{c}{$\alpha_{a}^{\sigma}$}
  & \multicolumn{1}{c}{$r_{\rm cut1}^\sigma$}
  & \multicolumn{1}{c}{$B_1^\sigma$}
  & \multicolumn{1}{c}{$B_2^\sigma$}
  & \multicolumn{1}{c}{$B_3^\sigma$}
  & \multicolumn{1}{c}{$B_4^\sigma$}
  & \multicolumn{1}{c}{$B_5^\sigma$}
  & \multicolumn{1}{c}{$B_6^\sigma$} \\
\midrule
\multirow{6}{*}{Mo}
 & 0 & 1.64185 & 1.14720 &  54.11237    & -34.45925  & -263.50099 &  132.40778   &   50.04099   & \dash \\
 & 0 & 1.64137 & 1.14720 & -200.69207   &  120.95773 &  886.66330  & -442.01817  & -166.92942   & \dash \\
 & 1 & 1.82347 & 0.99980 & -123.86769   &  400.72403 & -505.83259  &   12.12815  &  142.27439   & \dash \\
 & 1 & 1.82660 & 0.99980 &  241.18262   & -708.46248 &  858.79610  &  -12.57282  & -244.26709   & \dash \\
 & 2 & 1.95471 & 1.13300 &  330.97546   & -474.70460 &  162.00402  &   59.36305  &  -56.45340   & \dash \\
 & 2 & 1.93939 & 1.13300 & -504.58445   &  697.03148 & -197.46191  &  -90.11049  &   65.32075   & \dash \\
\midrule
\multirow{6}{*}{W}
 & 0 & 1.09400 & 1.48580 &  -23.14096   &   93.64875 &  -71.47339  &    3.73975  &    6.01571   & \dash \\
 & 0 & 1.09400 & 1.48580 &   87.99127   & -319.63486 &  243.76756  &  -17.49745  &  -18.66261   & \dash \\
 & 1 & 0.86000 & 1.60150 &   33.43550   &  -42.48764 &  -25.56101  &   33.92573  &   -7.63197   & \dash \\
 & 1 & 0.84500 & 1.60150 & -169.61662   &  177.23120 &  141.91974  & -161.87368  &   35.10767   & \dash \\
 & 2 & 0.59270 & 2.00560 &  -58.68214   &  120.24340 & -115.63716  &   45.62131  &   -5.81954   & \dash \\
 & 2 & 0.59325 & 2.00560 &   78.58040   & -155.01387 &  156.05920  &  -63.52092  &    8.23003   & \dash \\
\midrule
\multirow{4}{*}{S}
 & 0 & 3.52040 & 1.10780 &   -8.31868   &  -14.23766 &  -29.32160  &   46.92913  &   89.48958   &  0.00175    \\
 & 0 & 1.63490 & 1.10780 &   16.99407   &    6.64614 &   13.75514  & -122.78397  &  100.93015   & -25.15944   \\
 & 1 & 1.27480 & 1.27110 &   75.30803   &  -91.08992 & -207.45342  &  373.87021  & -189.94088   &  29.73608   \\
 & 1 & 1.25370 & 1.27110 &  -77.96879   &   68.27228 &  213.65044  & -346.61809  &  171.76343   & -26.99847   \\
\midrule
\multirow{4}{*}{Se}
 & 0 & 0.82593 & 1.38210 &   17.43200   &   99.75290 & -167.02200  &   78.29200  &  -29.71310   &  7.02445    \\
 & 0 & 0.82593 & 1.38210 &  -27.47330   & -126.48900 &  216.35600  & -101.75800  &   37.65720   & -8.74562    \\
 & 1 & 0.70950 & 1.41710 & -215.96800   & -198.51800 &  675.91600  & -425.03800  &  119.67800   & -15.40870   \\
 & 1 & 0.73858 & 1.41710 &  258.47700   &  231.88500 & -780.45400  &  479.87200  & -131.45800   &  16.60200   \\
\bottomrule
\end{tabular}%
\end{adjustbox}
\end{table*}

% ===================================================================
% TABLE (b) — alpha_a + rcut2 + C coefficients
% ===================================================================
\begin{table*}[t]
\centering
\footnotesize
\setlength{\tabcolsep}{4pt}
\renewcommand{\arraystretch}{1.0}
\caption{Best-fit parameters for the nonlocal pseudopotential beta functions of
Mo, W, S, and Se defined in Eq. (\ref{eq:combined}) for $r\geq R_s$. $\alpha_b^{\sigma}$, outer cutoff
$r_{\rm cut2}^\sigma$, and polynomial coefficients $C^\sigma_i$.
Dashes (--) indicate zero or unused.
All values rounded to five decimal places.}
\label{tab:beta_C}
\begin{adjustbox}{max width=\linewidth}
\begin{tabular}{@{} c c
    D{.}{.}{2.5} D{.}{.}{1.5}
    D{.}{.}{4.5} D{.}{.}{4.5} D{.}{.}{4.5}
    D{.}{.}{4.5} D{.}{.}{4.5} @{}}
\toprule
\textbf{Atom} & $l$
  & \multicolumn{1}{c}{$\alpha_b^\sigma$}
  & \multicolumn{1}{c}{$r_{\rm cut2}^\sigma$}
  & \multicolumn{1}{c}{$C_1^\sigma$}
  & \multicolumn{1}{c}{$C_2^\sigma$}
  & \multicolumn{1}{c}{$C_3^\sigma$}
  & \multicolumn{1}{c}{$C_4^\sigma$}
  & \multicolumn{1}{c}{$C_5^\sigma$} \\
\midrule
\multirow{6}{*}{Mo}
 & 0 & \dash & 1.49160 &   0.39153   &  -4.25928  &   12.43058   &  -24.94351   &  -31.86384   \\
 & 0 & \dash & 1.49160 &  -1.46871   &  14.19366  &  -33.83334   &    0.20104   &  116.33382   \\
 & 1 & 2.00582 & 1.56800 & -12.00351   &  39.94514  &  194.14493   & -324.26584   & -968.26338   \\
 & 1 & 2.10173 & 1.56800 &  21.59773   & -76.54876  & -308.74686   &  449.47136   & 1855.94778   \\
 & 2 & 6.73602 & 1.58780 &  -3.20471   & -28.26160  &   24.60902   & -394.09260   & 3098.78710   \\
 & 2 & 6.73602 & 1.58780 &   4.33450   &  32.95999  &  -46.20778   &  525.44577   & -3595.61275  \\
\midrule
\multirow{6}{*}{W}
 & 0 & 57.63840 & 2.24440 &   1.56915   & -11.93753  &   55.13174   & \dash & \dash \\
 & 0 & 61.04567 & 2.24440 &  -5.17588   &  40.04170  & -191.81477   & \dash & \dash \\
 & 1 & 31.16320 & 2.24440 &  -0.17052   &   0.55397  &   -1.74260   & \dash & \dash \\
 & 1 & 29.32013 & 2.24440 &   0.67791   &  -1.71934  &    4.97948   & \dash & \dash \\
 & 2 &  5.00000 & 2.03080 &   0.09628   &  -4.33311  & \dash & \dash & \dash \\
 & 2 &  5.00000 & 2.03080 &  -0.13873   &   6.24420  & \dash & \dash & \dash \\
\midrule
\multirow{4}{*}{S}
 & 0 & 7.95100 & 1.69450 &   2.92378   &   0.63661  &   -0.50191   &  134.34992   & -466.73081   \\
 & 0 & 8.12600 & 1.69450 &  -2.98341   &  -4.58559  &    9.41640   & -190.52041   &  612.59890   \\
 & 1 & 0.02790 & 1.69450 &  -0.42564   &  -7.75688  &  -50.29227   &  742.76841   & -1962.42117  \\
 & 1 & 2.86960 & 1.69450 &   0.18364   &   6.60448  &   50.78923   & -589.32071   & 1205.08075   \\
\midrule
\multirow{4}{*}{Se}
 & 0 & \dash & 1.91290 & -15.04280   &   8.03721  &  194.35800   & -570.45700   & -687.77800   \\
 & 0 & \dash & 1.91290 &  18.62740   &  -9.09707  & -245.25800   &  727.35600   &  801.13100   \\
 & 1 & \dash & 1.91290 &  27.19640   & -55.76360  & -255.02900   & 1648.30000   & -2464.52000  \\
 & 1 & \dash & 1.91290 & -30.38060   &  63.86260  &  277.91100   & -1818.37000  & 2721.99000   \\
\bottomrule
\end{tabular}%
\end{adjustbox}
\end{table*}

The parameters extracted from the DFT calculation for $E_{nn'}^{l}$ and $q_{nn'}^{l}$ are shown in Table~\ref{tab:Enn_qnn}.

\begin{table*}[t]
\centering
\footnotesize
\setlength{\tabcolsep}{4pt}
\renewcommand{\arraystretch}{1.0}
\caption{$E_{nn'}$ and $q_{nn'}$ parameters for different angular-momentum channels ($l$) in MoS$_2$, WS$_2$, MoSe$_2$, and WSe$_2$. All values are rounded uniformly to five decimal places.}
\label{tab:Enn_qnn}
\begin{adjustbox}{max width=\linewidth}
\begin{tabular}{@{} >{\centering\arraybackslash}c c c c *{2}{c c} *{2}{c c} *{2}{c c} *{2}{c c} @{}}
\toprule
\multirow{2}{*}{\textbf{Atom}} & \multirow{2}{*}{\textbf{n}} &\multirow{2}{*}{\textbf{n$'$}} & \multirow{2}{*}{\textbf{l}}
 & \multicolumn{2}{c}{\textbf{MoS$_2$}} 
 & \multicolumn{2}{c}{\textbf{WS$_2$}} 
 & \multicolumn{2}{c}{\textbf{MoSe$_2$}} 
 & \multicolumn{2}{c}{\textbf{WSe$_2$}} \\
\cmidrule(lr){5-6} \cmidrule(lr){7-8} \cmidrule(lr){9-10} \cmidrule(lr){11-12}
 &  &  & & $E_{nn'}$ & $q_{nn'}$ & $E_{nn'}$ & $q_{nn'}$ & $E_{nn'}$ & $q_{nn'}$ & $E_{nn'}$ & $q_{nn'}$ \\
\midrule
\multirow{9}{*}{\textbf{M}} 
 & 1 & 1 & 0 & 7.67651 & -0.28692 & 3.89237 & -0.84439 & 7.63563 & -0.28692 & 3.87563 & -0.84444 \\
 & 1 & 2 & 0 & 2.30705 & -0.08794 & 1.30435 & -0.27301 & 2.29415 & -0.08794 & 1.29911 & -0.27304 \\
 & 2 & 2 & 0 & 0.69774 & -0.02698 & 0.38972 & -0.09089 & 0.69413 & -0.02698 & 0.38801 & -0.09090 \\
 & 3 & 3 & 1 & 2.77239 & -0.14948 & -0.99014 & -0.19454 & 2.75330 & -0.14948 & -0.99442 & -0.19454 \\
 & 3 & 4 & 1 & 1.62184 & -0.08666 & -0.16468 & -0.06060 & 1.61070 & -0.08666 & -0.16600 & -0.06060 \\
 & 4 & 4 & 1 & 0.95008 & -0.05030 & -0.05511 & -0.01836 & 0.94357 & -0.05030 & -0.05550 & -0.01836 \\
 & 5 & 5 & 2 & 0.24333 & -0.00188 & -0.36676 &  0.02905 & 0.24380 & -0.00188 & -0.36603 &  0.02905 \\
 & 5 & 6 & 2 & 0.20148 &  0.00271 & -0.25923 &  0.02454 & 0.20230 &  0.00270 & -0.25860 &  0.02453 \\
 & 6 & 6 & 2 & 0.17693 &  0.00581 & -0.18049 &  0.02028 & 0.17797 &  0.00581 & -0.17995 &  0.02027 \\
\midrule
\multirow{6}{*}{\textbf{X}} 
 & 1 & 1 & 0 & 3.72921 & -0.22312 & 3.72825 & -0.22312 & 2.44423 & -0.31905 & 2.44201 & -0.31907 \\
 & 1 & 2 & 0 & 2.86369 & -0.13114 & 2.86309 & -0.13114 & 1.96962 & -0.24943 & 1.96789 & -0.24943 \\
 & 2 & 2 & 0 & 2.83706 & -0.06106 & 2.83678 & -0.06106 & 1.58884 & -0.19602 & 1.58748 & -0.19602 \\
 & 3 & 3 & 1 & 0.81553 & -0.04376 & 0.81535 & -0.04376 & 0.22857 & -0.06570 & 0.22809 & -0.06571 \\
 & 3 & 4 & 1 & 0.94965 & -0.04911 & 0.94944 & -0.04911 & 0.21243 & -0.05733 & 0.21202 & -0.05734 \\
 & 4 & 4 & 1 & 1.15065 & -0.05653 & 1.15041 & -0.05653 & 0.19696 & -0.05010 & 0.19660 & -0.05011 \\
\bottomrule
\end{tabular}%
\end{adjustbox}
\end{table*}

\section{Electronic Structure Calculations}

The all-electron Bloch state of band $\nu$ at in-plane wavevector $\mathbf{k}=(k_x,k_y)$ is constructed from a pseudo-wavefunction augmented by a projector term,
\begin{equation}
\psi_{\nu\mathbf{k}}(\boldsymbol{\rho},z) = \bigl(1 + \widehat{S}\bigr) \varphi_{\nu\mathbf{k}}(\boldsymbol{\rho},z)
\label{eq:bloch_sep}
\end{equation}
where $\boldsymbol{\rho}=(x,y)$ is the coordinate in the plane and $z$ is the coordinate outside the plane. The pseudo-wavefunction is expanded in a mixed B-spline and plane-wave basis,
\begin{equation}
\varphi_{\nu\mathbf{k}}(\boldsymbol{\rho},z) = \frac{1}{\sqrt{A}}\sum_{i=1}^{N}\sum_{\mathbf{G}} Z_{i\mathbf{G}}^{\nu,\mathbf{k}}\,B_i(z)\, e^{i(\mathbf{k}+\mathbf{G})\cdot\boldsymbol{\rho}}
\label{eq:bloch_basis}
\end{equation}
Here, $Z_{i\mathbf{G}}^{\nu,\mathbf{k}}$ are the expansion coefficients. The operator $\widehat{S}$ restores the all-electron character in the core regions via nonlocal projectors. Substituting \eqref{eq:bloch_sep} into the equation of all all-electron Schr\"{o}dinger
\begin{equation}
   \widehat{H}_{\mathrm{AE}} \psi_{\nu\mathbf{k}}(\boldsymbol{\rho},z) = E_{\nu}(\mathbf{k}) \psi_{\nu\mathbf{k}}(\boldsymbol{\rho},z) 
\end{equation}
and projecting onto the basis of \eqref{eq:bloch_basis} yields the generalized eigenvalue problem
\begin{equation}
\begin{split}
\sum_{i',\mathbf{G}'} \bigl\langle \mathbf{k}+\mathbf{G};B_i \bigm| \widehat{H} \bigm| \mathbf{k}+\mathbf{G}';B_{i'} \bigr\rangle Z_{i'\mathbf{G}'}^{\nu,\mathbf{k}} \\ = E_{\nu}(\mathbf{k}) \sum_{i',\mathbf{G}'} \bigl\langle \mathbf{k}+\mathbf{G};B_i \bigm| 1+\widehat{S} \bigm| \mathbf{k}+\mathbf{G}';B_{i'} \bigr\rangle Z_{i'\mathbf{G}'}^{\nu,\mathbf{k}}
\end{split}
\end{equation}
Here, $\widehat{H} = -\nabla^2 + \widehat{V}_{\rm loc} + \widehat{V}_{\rm nl}$ is the pseudo-Hamiltonian. The matrix elements for the kinetic-energy term are given by 
\begin{equation}
 \left\langle \mathbf{k}+\mathbf{G};B_i\middle|-\nabla^2  \middle|\mathbf{k}+\mathbf{G'};B_{i'} \right\rangle = [ K_{i,i'}+O_{i,i'} |\mathbf{k}+\mathbf{G}|^2 ] \delta_{\mathbf{G},\mathbf{G'}}
\end{equation}
with \( K_{i,i'} = \int dz (d/dz) B_i(z)  (d/dz) B_{i'} (z) \) and \( O_{i,i'} = \int dz  B_i(z)  B_{i'} (z) \). The matrix elements for the local potential term $\widehat{V}_{\rm loc} $ are
\begin{equation}
\begin{aligned}
&  \left\langle \mathbf{k}+\mathbf{G};B_i\middle|\widehat{V}_{\rm loc}  \middle|\mathbf{k}+\mathbf{G'};B_{i'} \right\rangle   \\
& = \frac{1}{A_{c}} \int dz  B_i(z)V^{\text{SEP}}_{\text{loc}}(z,\mathbf{G-G'})  B_{i'} (z) 
\end{aligned}
\end{equation}
where $V^{\text{SEP}}_{\text{loc}}(z,\mathbf{G-G'})$ is given in \eqref{eq:localpot}.
 
The matrix elements for the nonlocal pseudopotential term $\widehat{V}_{\rm nl}$ are
\begin{equation}\label{eq:matNLPP}
\begin{aligned}
& \left\langle \mathbf{K};B_{i} \middle| \widehat{V}_{nl} \middle| \mathbf{K}';B_{i'} \right\rangle \\
& = \sum_{\sigma lm,nn'} E_{lm}^{nn'} 
\left\langle \mathbf{K};B_{i} \middle| \beta_{lm}^{n\sigma} \right\rangle 
\left\langle \beta_{lm}^{n'\sigma} \middle| \mathbf{K}';B_{i'} \right\rangle 
e^{i(\mathbf{G}' - \mathbf{G}) \cdot \mathbf{\tau}_{\sigma}} \\
& = \frac{1}{A_{c}} \sum_{\sigma l,m \geq 0} \sum_{n,n'} E_{lm}^{nn'} 
\operatorname{Re}\left[ P_{lm}^{in\sigma}(\mathbf{K}) 
P_{lm}^{i'n'\sigma \ast}(\mathbf{K}') \right] 
e^{i(\mathbf{G}' - \mathbf{G}) \cdot \mathbf{\tau}_{\sigma}}
\end{aligned}
\end{equation}
where $\mathbf{K} = \mathbf{k} + \mathbf{G}$ and $\mathbf{K}' = \mathbf{k} + \mathbf{G}'$. The projection of the beta function in the mixed basis is given by
\begin{equation}\label{eq:matPlm}
\begin{split}
P_{lm}^{in\sigma}\left( \mathbf{K} \right) &= i^{l} \sqrt{A_{c}} \left\langle \mathbf{K}; B_{i} \middle| \beta_{lm}^{n\sigma} \right\rangle \\
&= \frac{1}{\sqrt{L_{c}}} \sum_{\mathbf{g}_{z}} \widetilde{B}_{i}\left( g_{z} \right) I_{l}(Q) Y_{lm}(\widehat{\mathbf{Q}})
\end{split}
\end{equation}
Here, \(\widetilde{B}_{i}(g_{z})\) is the Fourier transform of \(B_{i}(z)\), and $I_{l}(Q)$ is the radial integral.

By using the parameters \(E_{l}^{nn'}\) and \(q_{l}^{nn'}\) from Table~\ref{tab:Enn_qnn}, we can evaluate the matrix elements of \( \widehat{V}_{nl}\) and \(\widehat{S}\). To exploit the mirror symmetry ($z \to -z$) of monolayer TMDCs, we construct even and odd combinations of the B-splines:
\begin{equation}
B_i^{\pm}(z)
= \frac{1}{\sqrt{2}} \bigl[ B_i(z) \pm B_{N+1-i}(z) \bigr]
\label{eq:bsym}
\end{equation}
where $N$ is the total number of splines. This decouples the eigenvalue problem into two independent blocks, reducing the computational cost by approximately a factor of four.

\begin{figure*}[t!]
\centering
\includegraphics[
    width=\textwidth,
    trim=0.05cm 0.05cm 0.05cm 0.05cm,
    clip
]{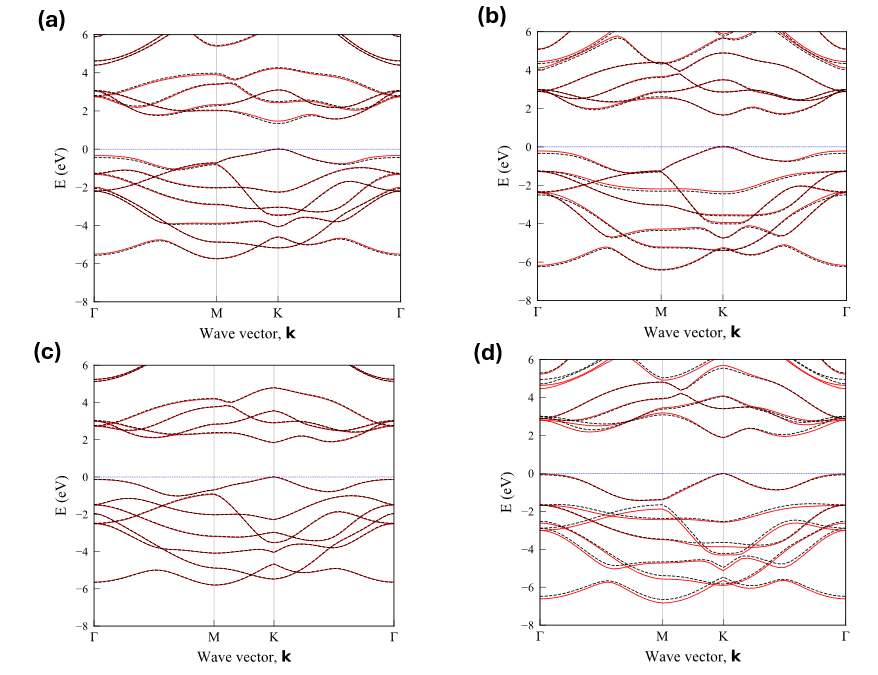}
\caption{Comparison of the band structures of four monolayer transition metal dichalcogenides calculated along the high-symmetry path $\Gamma$–M–K–$\Gamma$. Solid red lines denote results from the semi-empirical pseudopotential method (SEP), while dashed black lines indicate density-functional theory (DFT) band energies. The valence-band maximum (VBM) is aligned to zero energy, indicated by the horizontal blue dashed line. Subfigures show: (a) WSe\textsubscript{2}, (b) MoSe\textsubscript{2}, (c) MoS\textsubscript{2}, and (d) WS\textsubscript{2}. The comparison demonstrates that the SEP accurately reproduces the DFT-calculated band features.} 

\label{fig:band_comparison}
\end{figure*}

Figure~\ref{fig:band_comparison} presents a detailed comparison of the monolayer TMDC band structures along the high-symmetry path $\Gamma$--M--K--$\Gamma$, calculated using our SEP versus DFT approaches. The material exhibits a direct band gap at the \textbf{K} and \textbf{K'} points. The band structures obtained by the SEP method match the DFT results very well. We summarize the direct band gaps of the four monolayer TMDCs in Table~\ref{tab:bandgap_ml}.

\subsection{Transferability test: Bilayer WSe$_2$}
\label{sec:bilayer-transfer}

After validating the SEP framework for monolayer TMDCs, we tested its transferability by applying the monolayer-fitted atomic pseudopotentials to bilayer WSe$_2$ \emph{without} any additional refitting. This bilayer system is constructed by using the conventional 2H stacking geometry (Sect.~\ref{sec:atomic-basis}) As a reference, we compute the bilayer band structure using self-consistent DFT  ~\cite{ren2015mixed} as mentioned in Section 2, while the SEP band structure is obtained using the same atomic geometry. The interlayer separation adopted is 6.63 a.u. in both the SEP and DFT calculations. This comparison provides a test for the ability of the SEP developed for monolayer TMDC to calculate the near-band-gap electronic structures relevant for optical and transport properties for bilayer TMDC.

\begin{figure*}[t!]
\centering
\includegraphics[width=0.98\textwidth]{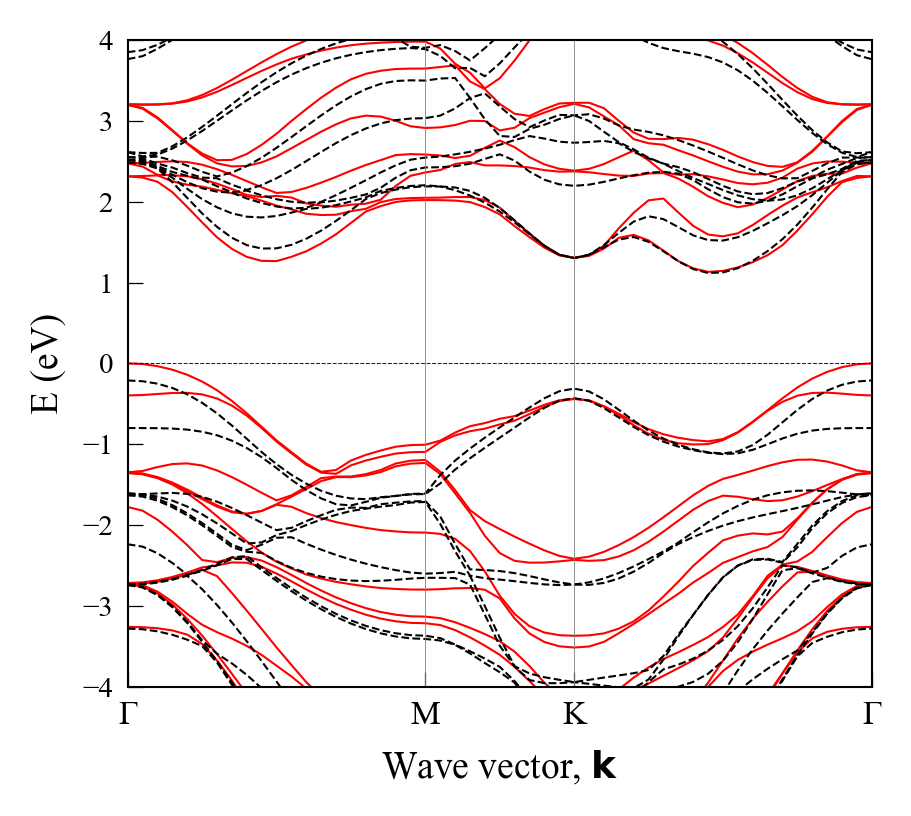}
\caption{Comparison of the electronic band structures of bilayer WSe$_2$ along the high-symmetry path $\Gamma$--M--K--$\Gamma$. Solid red lines are obtained from the semi-empirical pseudopotential (SEP) method, and dashed black lines are from self-consistent DFT calculations.  The horizontal blue dashed line indicates the valence-band maximum (VBM) of SEP results. The band structures obtained via the DFT calculation are rigidly shifted so the conduction-band minimum (CBM) is aligned with the corresponding SEP result for comparison.}
\label{fig:band_comparison_Bl_WSe2}
\end{figure*}

\begin{table}[H]
\centering
\caption{Lattice constant ($a$), distance between two chalcogen atoms ($2\tau_c$), and direct band gap ($E_g$) at the K point for monolayer 
transition metal dichalcogenides (TMDCs) obtained from self-consistent DFT 
and the semi-empirical pseudopotential (SEP) method. $\Delta E_g$ 
denotes the difference between SEP and DFT band gap values.}
\label{tab:bandgap_ml}
\small
\begin{tabular}{lccccc}
\hline
Material & $a$ (\AA) & $2\tau_c$ (\AA) & $E_g^{\mathrm{DFT}}$ (eV) & $E_g^{\mathrm{SEP}}$ (eV) & $\Delta E_g$ (eV) \\
\hline
WSe$_2$  & 3.315 & 3.367& 1.65 & 1.68 & 0.03  \\
MoSe$_2$ & 3.299 & 3.338 & 1.35 & 1.46 & 0.11 \\
MoS$_2$  & 3.160 & 3.172 & 1.84 & 1.85 & 0.01  \\
WS$_2$   & 3.186 & 3.142 & 1.90 & 1.87 & 0.03  \\
\hline
\end{tabular}
\end{table}

Figure~\ref{fig:band_comparison_Bl_WSe2} shows the band structures of bilayer WSe$_2$ obtained by using SEP obtained in this work compared to those obtained by using DFT. Overall, SEP reproduces the DFT dispersions near the valence- and conduction-band edges with good agreement along the paths M-K and K-Q, capturing the energy positions and curvatures of the frontier bands. Deviations become more noticeable for deeper valence states and for bands with stronger interlayer hybridization, particularly near $\Gamma$, which is expected because the present SEP is fitted at the monolayer level and does not take into account stacking-induced charge redistribution. Despite these discrepancies, the agreement for the bands near the band gap at the K point demonstrates that the monolayer-derived SEP remains reliable in the presence of interlayer coupling, supporting its use for efficient electronic-structure studies of layered TMDCs. It should be noted that the main qualitative difference in monolayer and bilayer WSe$_2$ is the conduction-band (CB) energy at the Q point. The CBM is at the  K point in monolayer WSe$_2$, making it a direct band-gap material. In contrast, the CBM is at the Q point in bilayer WSe$_2$, making it an indirect band-gap material. This important feature is reproduced by using the SEP obtained here. The noticeable deviations between the SEP and DFT results in the deeper valence bands may be remedied in the future by adding a correction potential in the interfacial region.

\subsection{Computational efficiency and timing comparison}
\label{sec:timing}

We now compare the computational cost of the semi-empirical pseudopotential (SEP) approach with that of conventional self-consistent DFT calculations for monolayer WSe$_2$. In standard DFT workflows, we first perform an iterative procedure to obtain the self-consistent potential (SCP), which requires repeated construction of the Hamiltonian matrix and diagonalization. We then adopt the resulting SCP to calculate the band structure. In the SEP method, we only need to construct the effective single-particle Hamiltonian once and evaluate the band structure. 

As an illustrative comparison, we report representative wall-clock runtime obtained on the same computational platform using the Fortran platform. For monolayer WSe$_2$, the self-consistent DFT workflow—converged to a variation in the effective potential below $10^{-7}$~Ry—requires a total run time of approximately 552 seconds for the complete calculation of SCP and band-structure. In contrast, the corresponding SEP calculation for the same system completes in about 80 seconds, highlighting the substantial reduction in computational cost achieved by eliminating the self-consistent computation.

\subsection*{Relation to real-space electronic-structure methods}
It is useful to clarify the scope of the present semi-empirical pseudopotential
(SEP) approach relative to modern real-space electronic-structure methods.
Real-space formulations of density functional theory, such as finite-difference
and localized-basis implementations, offer great flexibility in treating various
boundary conditions and have demonstrated excellent scalability to very large
systems through iterative diagonalization techniques ~\cite{Natan2008PRB,Dogan2023JCP}. 
These methods are particularly effective for obtaining self-consistent total energies 
and charge densities in systems containing tens or hundreds of thousands of atoms.

The SEP method developed here is not intended to replace such large-scale
self-consistent real-space DFT solvers. Instead, it targets a complementary
regime in which an effective single-particle Hamiltonian is constructed by fitting DFT results. Afterwards, the electronic states
can be calculated efficiently without self-consistent iterations. This
makes SEP well suited for applications that require dense Brillouin-zone sampling or rapid exploration of stacking
configurations in low-dimensional materials. Accordingly, SEP and real-space self-consistent methods address different
computational objectives and should be regarded as complementary tools
within the broader electronic-structure landscape. Since these semi-empirical potentials are mostly in simple analytical forms, they can also be used in a semi-empirical real-space approach by using tight-binding orbitals as the basis. The matrix elements for these SEPs within the tight-binding basis can be easily constructed.

\section{Conclusion}
We have developed a semi-empirical pseudopotential (SEP) method for monolayer TMDCs that can reproduce the band structures calculated by using DFT with decent accuracy. The SEP code is efficient, and straightforward to implement. By eliminating the self-consistent procedure, the SEP approach substantially reduces the computational cost of band-structure calculations compared to self-consistent DFT, while retaining good quantitative agreement for representative TMDC systems. The computational efficiency makes the method well suited for rapid electronic-structure evaluation and exploratory modeling of TMDC-based nanostructures and optoelectronic devices.

The SEP developed here for monolayer TMDCs also provides a good starting point for constructing SEPs for multilayer TMDCs. We have demonstrated that the SEP of monolayer WSe$_2$ can be used to calculate the band structure of bilayer WSe$_2$ with qualitatively correct features around the K and Q points. To obtain more accurate band
structures throughout the entire Brillouin zone, additional work is still needed to capture the effect caused by charge redistribution near the interface between layers. The spin-orbit coupling also needs to be incorporated in the future to describe the spin-orbit splitting in these materials.

\section{Conflict of Interest}
The authors have no conflicts to disclose.

\section{Data Availability}
The data supporting the findings of this study, including the SEP Fortran implementation and representative input files used in this work, will be made publicly available in a public repository upon publication, enabling full reproducibility of the reported results.

\section{acknowledgment}
The authors gratefully acknowledge Dr. Tung-Han Hsieh (RCAS, Academia Sinica) for technical support. Work funded by Taiwan National Science and Technology Council under grant numbers: NSTC 112-2112-M-001-054-MY2 and  NSTC 114-2112-M-006-030.

%% References
\bibliography{references} 

@PREAMBLE{
 "\providecommand{\noopsort}[1]{}" 
 # "\providecommand{\singleletter}[1]{#1}%" 
}

@article{novoselov2004electric,
  title={Electric field effect in atomically thin carbon films},
  author={Novoselov, Kostya S. and Geim, Andre K. and Morozov, Sergei V. and Jiang, De-eng and Zhang, Yanshui and Dubonos, Sergey V. and Grigorieva, Irina V. and Firsov, Alexandr A.},
  journal={Science},
  volume={306},
  number={5696},
  pages={666--669},
  year={2004},
  doi={10.1126/science.1102896},
  url={https://doi.org/10.1126/science.1102896}
}

@article{mak2010atomically,
  title={Atomically thin {MoS$_2$}: a new direct-gap semiconductor},
  author={Mak, Kin Fai and Lee, Changgu and Hone, James and Shan, Jie and Heinz, Tony F.},
  journal={Physical Review Letters},
  volume={105},
  number={13},
  pages={136805},
  year={2010},
  doi={10.1103/PhysRevLett.105.136805},
  url={https://doi.org/10.1103/PhysRevLett.105.136805}
}

@article{wang2012electronics,
  title={Electronics and optoelectronics of two-dimensional transition metal dichalcogenides},
  author={Wang, Qing Hua and Kalantar-Zadeh, Kourosh and Kis, Andras and Coleman, Jonathan N. and Strano, Michael S.},
  journal={Nature Nanotechnology},
  volume={7},
  number={11},
  pages={699--712},
  year={2012},
  doi={10.1038/nnano.2012.193},
  url={https://doi.org/10.1038/nnano.2012.193}
}

@article{qi2016superconductivity,
  title={Superconductivity in {Weyl} semimetal candidate {MoTe$_2$}},
  author={Qi, Yanpeng and Naumov, Pavel G. and Ali, Mazhar N. and Rajamathi, Catherine R. and Schnelle, Walter and Barkalov, Oleg and Hanfland, Michael and Wu, Shu-Chun and Shekhar, Chandra and Sun, Yan and B{\"u}chner, Bernd and Greenblatt, Martha and Felser, Claudia and Parkin, Stuart S. P.},
  journal={Nature Communications},
  volume={7},
  number={1},
  pages={11038},
  year={2016},
  doi={10.1038/ncomms11038},
  url={https://doi.org/10.1038/ncomms11038}
}

@article{li2018metallic,
  title={Metallic transition-metal dichalcogenide nanocatalysts for energy conversion},
  author={Li, Haoyi and Jia, Xiaofan and Zhang, Qi and Wang, Xun},
  journal={Chem},
  volume={4},
  number={7},
  pages={1510--1537},
  year={2018},
  doi={10.1016/j.chempr.2018.03.012},
  url={https://doi.org/10.1016/j.chempr.2018.03.012}
}

@article{wu2019topological,
  title={Topological insulators in twisted transition metal dichalcogenide homobilayers},
  author={Wu, Fengcheng and Lovorn, Timothy and Tutuc, Emanuel and Martin, Ivar and MacDonald, A. H.},
  journal={Physical Review Letters},
  volume={122},
  number={8},
  pages={086402},
  year={2019},
  doi={10.1103/PhysRevLett.122.086402},
  url={https://doi.org/10.1103/PhysRevLett.122.086402}
}

@article{liu2014optical,
  title={Optical properties of monolayer transition metal dichalcogenides probed by spectroscopic ellipsometry},
  author={Liu, Hsiang-Lin and Shen, Chih-Chiang and Su, Sheng-Han and Hsu, Chang-Lung and Li, Ming-Yang and Li, Lain-Jong},
  journal={Applied Physics Letters},
  volume={105},
  number={20},
  pages={201105},
  year={2014},
  doi={10.1063/1.4901836},
  url={https://doi.org/10.1063/1.4901836}
}

@article{johari2012tuning,
  title={Tuning the electronic properties of semiconducting transition metal dichalcogenides by applying mechanical strains},
  author={Johari, Priya and Shenoy, Vivek B.},
  journal={ACS Nano},
  volume={6},
  number={6},
  pages={5449--5456},
  year={2012},
  doi={10.1021/nn301320r},
  url={https://pubs.acs.org/doi/10.1021/nn301320r}
}

@article{shi2018mechanical,
  title={Mechanical and electronic properties of {Janus} monolayer transition metal dichalcogenides},
  author={Shi, Wenwu and Wang, Zhiguo},
  journal={Journal of Physics: Condensed Matter},
  volume={30},
  number={21},
  pages={215301},
  year={2018},
  doi={10.1088/1361-648X/aabd59},
  url={https://iopscience.iop.org/article/10.1088/1361-648X/aabd59}
}

@article{chen2020environmental,
  title={Environmental analysis with {2D} transition-metal dichalcogenide-based field-effect transistors},
  author={Chen, Xiaoyan and Liu, Chengbin and Mao, Shun},
  journal={Nano-Micro Letters},
  volume={12},
  pages={1--24},
  year={2020},
  doi={10.1007/s40820-020-00438-w},
  url={https://doi.org/10.1007/s40820-020-00438-w}
}

@article{li2014metal,
  title={Metal-semiconductor barrier modulation for high photoresponse in transition metal dichalcogenide field effect transistors},
  author={Li, Hua-Min and Lee, Dae-Yeong and Choi, Min Sup and Qu, Deshun and Liu, Xiaochi and Ra, Chang-Ho and Yoo, Won Jong},
  journal={Scientific Reports},
  volume={4},
  number={1},
  pages={4041},
  year={2014},
  doi={10.1038/srep04041},
  url={https://doi.org/10.1038/srep04041}
}

@article{ko2016improvement,
  title={Improvement of gas-sensing performance of large-area tungsten disulfide nanosheets by surface functionalization},
  author={Ko, Kyung Yong and Song, Jeong-Gyu and Kim, Youngjun and Choi, Taejin and Shin, Sera and Lee, Chang Wan and Lee, Kyounghoon and Koo, Jahyun and Lee, Hoonkyung and Kim, Jongbaeg and Park, Seong-Jun and Lee, Zonghoon and Park, Hyung-Ho and Lee, Takhee},
  journal={ACS Nano},
  volume={10},
  number={10},
  pages={9287--9296},
  year={2016},
  doi={10.1021/acsnano.6b03631},
  url={https://doi.org/10.1021/acsnano.6b03631}
}

@article{paudel2023semi,
  title={Semi-empirical pseudopotential method for graphene and graphene nanoribbons},
  author={Paudel, Raj Kumar and Ren, Chung-Yuan and Chang, Yia-Chung},
  journal={Nanomaterials},
  volume={13},
  number={14},
  pages={2066},
  year={2023},
  doi={10.3390/nano13142066},
  url={https://doi.org/10.3390/nano13142066}
}

@article{hohenberg1964inhomogeneous,
  title={Inhomogeneous electron gas},
  author={Hohenberg, Pierre and Kohn, Walter},
  journal={Physical Review},
  volume={136},
  number={3B},
  pages={B864},
  year={1964},
  doi={10.1103/PhysRev.136.B864},
  url={https://doi.org/10.1103/PhysRev.136.B864}
}

@article{qiu2013optical,
  title={Optical spectrum of {MoS$_2$}: many-body effects and diversity of exciton states},
  author={Qiu, Diana Y. and da Jornada, Felipe H. and Louie, Steven G.},
  journal={Physical Review Letters},
  volume={111},
  number={21},
  pages={216805},
  year={2013},
  doi={10.1103/PhysRevLett.111.216805},
  url={https://doi.org/10.1103/PhysRevLett.111.216805}
}

@article{li1994electronic,
  title={Electronic structures of {As/Si(001) 2$\times$1} and {Sb/Si(001) 2$\times$1} surfaces},
  author={Li, Guangwei and Chang, Yia-Chung},
  journal={Physical Review B},
  volume={50},
  number={12},
  pages={8675},
  year={1994},
  doi={10.1103/PhysRevB.50.8675},
  url={https://doi.org/10.1103/PhysRevB.50.8675}
}

@article{chang1996planar,
  title={Planar-basis pseudopotential method and planar {Wannier} functions for surfaces and heterostructures},
  author={Chang, Yia-Chung and Li, Guangwei},
  journal={Computer Physics Communications},
  volume={95},
  number={2-3},
  pages={158--170},
  year={1996},
  doi={10.1016/0010-4655(96)00056-2},
  url={https://doi.org/10.1016/0010-4655(96)00056-2}
}

@article{ren2015mixed,
  title={A mixed basis density functional approach for low dimensional systems with {B}-splines},
  author={Ren, Chung-Yuan and Hsue, Chen-Shiung and Chang, Yia-Chung},
  journal={Computer Physics Communications},
  volume={188},
  pages={94--102},
  year={2015},
  doi={10.1016/j.cpc.2014.11.013},
  url={https://doi.org/10.1016/j.cpc.2014.11.013}
}

@article{ren2022density,
  title={Density functional calculations of atomic structure, charging effect, and static dielectric constant of two-dimensional systems based on {B}-splines},
  author={Ren, Chung-Yuan and Chang, Yia-Chung},
  journal={Physica E: Low-dimensional Systems and Nanostructures},
  volume={140},
  pages={115203},
  year={2022},
  doi={10.1016/j.physe.2022.115203},
  url={https://doi.org/10.1016/j.physe.2022.115203}
}

@article{bachau2001applications,
  title={Applications of {B}-splines in atomic and molecular physics},
  author={Bachau, Henri and Cormier, Eric and Decleva, Piero and Hansen, J. E. and Mart{\'\i}n, Fernando},
  journal={Reports on Progress in Physics},
  volume={64},
  number={12},
  pages={1815},
  year={2001},
  doi={10.1088/0034-4885/64/12/205},
  url={https://iopscience.iop.org/article/10.1088/0034-4885/64/12/205}
}

@article{frigo2005design,
  title={The design and implementation of {FFTW3}},
  author={Frigo, Matteo and Johnson, Steven G.},
  journal={Proceedings of the IEEE},
  volume={93},
  number={2},
  pages={216--231},
  year={2005},
  doi={10.1109/JPROC.2004.840301},
  url={http://dx.doi.org/10.1109/JPROC.2004.840301}
}

@article{garrity2014pseudopotentials,
  title={Pseudopotentials for high-throughput {DFT} calculations},
  author={Garrity, Kevin F. and Bennett, Joseph W. and Rabe, Karin M. and Vanderbilt, David},
  journal={Computational Materials Science},
  volume={81},
  pages={446--452},
  year={2014},
  doi={10.1016/j.commatsci.2013.08.053},
  url={https://doi.org/10.1016/j.commatsci.2013.08.053}
}

@article{cohen1966band,
  title={Band structures and pseudopotential form factors for fourteen semiconductors of the diamond and zinc-blende structures},
  author={Cohen, Marvin L. and Bergstresser, T. K.},
  journal={Physical Review},
  volume={141},
  number={2},
  pages={789},
  year={1966},
  doi={10.1103/PhysRev.141.789},
  url={https://doi.org/10.1103/PhysRev.141.789}
}

@article{chelikowsky1976nonlocal,
  title={Nonlocal pseudopotential calculations for the electronic structure of eleven diamond and zinc-blende semiconductors},
  author={Chelikowsky, James R. and Cohen, Marvin L.},
  journal={Physical Review B},
  volume={14},
  number={2},
  pages={556},
  year={1976},
  doi={10.1103/PhysRevB.14.556},
  url={https://doi.org/10.1103/PhysRevB.14.556}
}

@article{vanderbilt1990soft,
  title={Soft self-consistent pseudopotentials in a generalized eigenvalue formalism},
  author={Vanderbilt, David},
  journal={Physical Review B},
  volume={41},
  number={11},
  pages={7892},
  year={1990},
  doi={10.1103/PhysRevB.41.7892},
  url={https://doi.org/10.1103/PhysRevB.41.7892}
}

@book{deBoor1987practical,
  title={A practical guide to splines},
  author={de Boor, Carl},
  year={1978},
  publisher={Springer-Verlag},
  address={New York},
  edition={Revised},
  note={ISBN-10: 0387953663},
  url={https://link.springer.com/book/10.1007/978-1-4612-6333-3}
}

@book{martin2020electronic,
  title={Electronic structure: Basic theory and practical methods},
  author={Martin, Richard M.},
  year={2020},
  publisher={Cambridge University Press},
  edition={2nd},
  isbn={9781108429900},
  url={https://www.cambridge.org/us/academic/subjects/physics/condensed-matter-physics-nanoscience-and-mesoscopic-physics/electronic-structure-basic-theory-and-practical-methods-2nd-edition}
}

@article{perdew1996generalized,
  title={Generalized gradient approximation made simple},
  author={Perdew, John P. and Burke, Kieron and Ernzerhof, Matthias},
  journal={Physical Review Letters},
  volume={77},
  number={18},
  pages={3865},
  year={1996},
  doi={10.1103/PhysRevLett.77.3865},
  url={https://doi.org/10.1103/PhysRevLett.77.3865}
}

@article{herring1940new,
  title={A new method for calculating wave functions in crystals},
  author={Herring, Conyers},
  journal={Physical Review},
  volume={57},
  number={12},
  pages={1169},
  year={1940},
  doi={10.1103/PhysRev.57.1169},
  url={https://doi.org/10.1103/PhysRev.57.1169}
}

@article{phillips1959new,
  title={New method for calculating wave functions in crystals and molecules},
  author={Phillips, James C. and Kleinman, Leonard},
  journal={Physical Review},
  volume={116},
  number={2},
  pages={287},
  year={1959},
  doi={10.1103/PhysRev.116.287},
  url={https://doi.org/10.1103/PhysRev.116.287}
}

@article{kleinman1982efficacious,
  title={Efficacious form for model pseudopotentials},
  author={Kleinman, Leonard and Bylander, D. M.},
  journal={Physical Review Letters},
  volume={48},
  number={20},
  pages={1425},
  year={1982},
  doi={10.1103/PhysRevLett.48.1425},
  url={https://doi.org/10.1103/PhysRevLett.48.1425}
}

@article{wang1996pseudopotential,
  title={Pseudopotential calculations of nanoscale {CdSe} quantum dots},
  author={Wang, Lin-Wang and Zunger, Alex},
  journal={Physical Review B},
  volume={53},
  number={15},
  pages={9579},
  year={1996},
  doi={10.1103/PhysRevB.53.9579},
  url={https://doi.org/10.1103/PhysRevB.53.9579}
}

@article{wang1995local,
  title={Local-density-derived semiempirical pseudopotentials},
  author={Wang, Lin-Wang and Zunger, Alex},
  journal={Physical Review B},
  volume={51},
  number={24},
  pages={17398},
  year={1995},
  doi={10.1103/PhysRevB.51.17398},
  url={https://doi.org/10.1103/PhysRevB.51.17398}
}

@article{fu1997local,
  title={Local-density-derived semiempirical nonlocal pseudopotentials for {InP} with applications to large quantum dots},
  author={Fu, Huaxiang and Zunger, Alex},
  journal={Physical Review B},
  volume={55},
  number={3},
  pages={1642},
  year={1997},
  doi={10.1103/PhysRevB.55.1642},
  url={https://doi.org/10.1103/PhysRevB.55.1642}
}

@article{bester2008electronic,
  title={Electronic excitations in nanostructures: an empirical pseudopotential based approach},
  author={Bester, Gabriel},
  journal={Journal of Physics: Condensed Matter},
  volume={21},
  number={2},
  pages={023202},
  year={2008},
  doi={10.1088/0953-8984/21/2/023202},
  url={https://doi.org/10.1088/0953-8984/21/2/023202}
}

@article{molina2012semiempirical,
  title={Semiempirical pseudopotential approach for nitride-based nanostructures and ab initio based passivation of free surfaces},
  author={Molina-S{\'a}nchez, Alejandro and Garc{\'\i}a-Crist{\'o}bal, Alberto and Bester, Gabriel},
  journal={Physical Review B},
  volume={86},
  number={20},
  pages={205430},
  year={2012},
  doi={10.1103/PhysRevB.86.205430},
  url={https://doi.org/10.1103/PhysRevB.86.205430}
}

@article{ren2023density,
  title={Density functional theory for buckyballs within symmetrized icosahedral basis},
  author={Ren, Chung-Yuan and Paudel, Raj Kumar and Chang, Yia-Chung},
  journal={Nanomaterials},
  volume={13},
  number={13},
  pages={1912},
  year={2023},
  doi={10.3390/nano13131912},
  url={https://doi.org/10.3390/nano13131912}
}

@techreport{shewchuk1994introduction,
  title={An introduction to the conjugate gradient method without the agonizing pain},
  author={Shewchuk, Jonathan Richard},
  year={1994},
  institution={Carnegie-Mellon University, Department of Computer Science},
  address={Pittsburgh, PA},
  number={CMU-CS-94-138},
  url={https://www.cs.cmu.edu/~quake-papers/painless-conjugate-gradient.pdf}
}

@article{kohn1965self,
  title={Self-consistent equations including exchange and correlation effects},
  author={Kohn, Walter and Sham, Lu Jeu},
  journal={Physical Review},
  volume={140},
  number={4A},
  pages={A1133},
  year={1965},
  doi={10.1103/PhysRev.140.A1133},
  url={https://doi.org/10.1103/PhysRev.140.A1133}
}

@article{zhu2011giant,
  title={Giant spin-orbit-induced spin splitting in two-dimensional transition-metal dichalcogenide semiconductors},
  author={Zhu, Zhiyong Y. and Cheng, Yingchun C. and Schwingenschl{\"o}gl, Udo},
  journal={Physical Review B},
  volume={84},
  number={15},
  pages={153402},
  year={2011},
  doi={10.1103/PhysRevB.84.153402},
  url={https://doi.org/10.1103/PhysRevB.84.153402}
}

@article{pandey1974nonlocal,
  title={Nonlocal pseudopotentials for {Ge} and {GaAs}},
  author={Pandey, K. C. and Phillips, J. C.},
  journal={Physical Review B},
  volume={9},
  number={4},
  pages={1552},
  year={1974},
  doi={10.1103/PhysRevB.9.1552},
  url={https://doi.org/10.1103/PhysRevB.9.1552}
}

@article{kresse1996efficiency,
  title={Efficiency of ab-initio total energy calculations for metals and semiconductors using a plane-wave basis set},
  author={Kresse, Georg and Furthm{\"u}ller, J{\"u}rgen},
  journal={Computational materials science},
  volume={6},
  number={1},
  pages={15--50},
  year={1996},
  publisher={Elsevier}
}

@article{mattheiss1973band,
  title={Band structures of transition-metal-dichalcogenide layer compounds},
  author={Mattheiss, L. F.},
  journal={Physical Review B},
  volume={8},
  number={8},
  pages={3719},
  year={1973},
  doi={10.1103/PhysRevB.8.3719},
  url={https://doi.org/10.1103/PhysRevB.8.3719}
}

@article{yun2022escalating,
  title={Escalating ferromagnetic order via {Se}-vacancies near vanadium in {WSe2} monolayers},
  author={Yun, Seok Joon and Cho, Byeong Wook and Thapa, Dinesh and Yang, Dae Hee and Kim, Yong In and Jin, Jeong Won and Yang, Sang-Hyeok and Nguyen, Tuan Dung and Kim, Young-Min and Kim, Ki Kang and Rho, Heesun},
  journal={Advanced Materials},
  volume={34},
  number={10},
  pages={2106551},
  year={2022},
  doi={10.1002/adma.202106551},
  url={https://doi.org/10.1002/adma.202106551}
}

@article{QE-2009,
  title={{QUANTUM ESPRESSO}: a modular and open-source software project for quantum simulations of materials},
  author={Giannozzi, Paolo and Baroni, Stefano and Bonini, Nicola and Calandra, Matteo and Car, Roberto and Cavazzoni, Carlo and Ceresoli, Davide and Chiarotti, Guido L. and Cococcioni, Matteo and Dabo, Ismaila and Dal Corso, Andrea and de Gironcoli, Stefano and Fabris, Stefano and Fratesi, Guido and Gebauer, Ralph and Gerstmann, Uwe and Gougoussis, Christos and Kokalj, Anton and Lazzeri, Michele and Martin-Samos, Layla and Marzari, Nicola and Mauri, Francesco and Mazzarello, Riccardo and Paolini, Stefano and Pasquarello, Alfredo and Paulatto, Lorenzo and Sbraccia, Carlo and Scandolo, Sandro and Sclauzero, Gabriele and Seitsonen, Ari P. and Smogunov, Alexander and Umari, Paolo and Wentzcovitch, Renata M.},
  journal={Journal of Physics: Condensed Matter},
  volume={21},
  number={39},
  pages={395502},
  year={2009},
  doi={10.1088/0953-8984/21/39/395502},
  url={https://dx.doi.org/10.1088/0953-8984/21/39/395502}
}

@article{kim2024transferable,
  title={Transferable empirical pseudopotenials from machine learning},
  author={Kim, Rokyeon and Son, Young-Woo},
  journal={Physical Review B},
  volume={109},
  number={4},
  pages={045153},
  year={2024},
  doi={10.1103/PhysRevB.109.045153},
  url={https://dx.doi.org/10.1103/PhysRevB.109.045153}
}

@misc{lin2025deep,
  title={Deep-learning atomistic pseudopotential model for nanomaterials},
  author={Lin, Kailai and Coley-O'Rourke, Matthew J. and Rabani, Eran},
  year={2025},
  eprint={2505.09846},
  archivePrefix={arXiv},
  primaryClass={cond-mat.mtrl-sci}
}

@article{Natan2008PRB,
  title = {Real-space pseudopotential method for first principles calculations of general periodic and partially periodic systems},
  author = {Natan, Amir and Benjamini, Ayelet and Naveh, Doron and Kronik, Leeor and Tiago, Murilo L. and Beckman, Scott P. and Chelikowsky, James R.},
  journal = {Phys. Rev. B},
  volume = {78},
  issue = {7},
  pages = {075109},
  numpages = {10},
  year = {2008},
  month = {Aug},
  publisher = {American Physical Society},
  doi = {10.1103/PhysRevB.78.075109},
  url = {https://link.aps.org/doi/10.1103/PhysRevB.78.075109}
}

@article{Dogan2023JCP,
    author = {Dogan, Mehmet and Liou, Kai-Hsin and Chelikowsky, James R.},
    title = {Real-space solution to the electronic structure problem for nearly a million electrons},
    journal = {The Journal of Chemical Physics},
    volume = {158},
    number = {24},
    pages = {244114},
    year = {2023},
    month = {06},
    issn = {0021-9606},
    doi = {10.1063/5.0150864},
    url = {https://doi.org/10.1063/5.0150864},
}

@article{gorelik20213d,
  title={3D electron diffraction of mono-and few-layer MoS$_2$},
  author={Gorelik, Tatiana E and Nergis, Berkin and Schoener, Tobias and Koester, Janis and Kaiser, Ute},
  journal={Micron},
  volume={146},
  pages={103071},
  year={2021},
  publisher={Elsevier}
}

\end{document}